\documentclass{SciPost}

\hypersetup{
    colorlinks,
    linkcolor={red!50!black},
    citecolor={blue!50!black},
    urlcolor={blue!80!black}
}

\usepackage{tikz-cd}
\usepackage{braket}
\usepackage{bm}
\usepackage{dsfont}

\usepackage[bitstream-charter]{mathdesign}
\DeclareSymbolFont{usualmathcal}{OMS}{cmsy}{m}{n}
\DeclareSymbolFontAlphabet{\mathcal}{usualmathcal}

\fancypagestyle{SPstyle}{
\fancyhf{}
\lhead{\colorbox{scipostblue}{\bf \color{white} ~SciPost Physics }}
\rhead{{\bf \color{scipostdeepblue} ~Submission }}

\fancyfoot[C]{\textbf{\thepage}}
}

\def \q{\mathbf{q}}
\newcommand{\1}{\mathds{1}}
\newcommand{\ubar}[1]{\underline{#1}}
\newcommand{\trace}{\mathrm{Tr}}
\newcommand{\tr}{\mathrm{Tr}}
\newcommand{\N}{\mathcal{N}}
\newcommand{\E}{\mathcal{E}}

\begin{document}

\pagestyle{SPstyle}

\begin{center}{\Large \textbf{\color{scipostdeepblue}{
Local correlations in long-range dual-unitary kicked Hamiltonian chains\\
}}}\end{center}

\begin{center}\textbf{
Vladimir Al. Osipov\textsuperscript{1,2$\star$},
Marc C. Spyra\textsuperscript{3} 
Jana C. Schumann\textsuperscript{3} 
Thomas Guhr\textsuperscript{3} and
Boris Gutkin\textsuperscript{4$\dagger$}
}\end{center}

\begin{center}
{\bf 1} Institute for Advanced Study in Mathematics, Harbin Institute of Technology, West Da Zhi Str. 92, 150001 Harbin, China
\\
{\bf 2} Suzhou Research Institute, Harbin Institute of Technology, South Guandu Road 500, 215104 Suzhou, China
\\
{\bf 3} Duisburg-Essen University, Lotharstr. 1, 47048 Duisburg, Germany
\\
{\bf 4} Department of Applied Mathematics, Holon Institute of Technology, 58102 Holon, Israel
\\[\baselineskip]
$\star$ \href{mailto:Vladimir.Al.Osipov@gmail.com}{\small Vladimir.Al.Osipov@gmail.com}\,,\quad
$\dagger$ \href{mailto:borisgu@hit.ac.il}{\small borisgu@hit.ac.il}
\end{center}

\section*{\color{scipostdeepblue}{Abstract}}
\textbf{\boldmath{%
Many-body Floquet models with exact space--time duality, such as the kicked Ising chain, provide paradigmatic examples of dual-unitary dynamics. The requirement of exact space--time duality is, however, highly restrictive and, in its standard formulation, permits only nearest-neighbor interactions. We demonstrate that the latter constraint can be avoided, still resulting in exactly solvable models. In particular, we construct a broad class of dual-unitary kicked spin chains with long-range interactions. We show that, for an interaction range $r$, local two-point connected correlation functions propagate along the edges of the light cone, $|n|=r|t|$, and can be evaluated analytically for local observables. We illustrate our general statement by analytic calculation of the connected correlation function for a kicked Ising chain with next-to-next-nearest-neighbor interactions. 
}}

\vspace{\baselineskip}





\date{\today}

\section{Introduction}
 
Many-body quantum systems pose formidable challenges for both
analytical and numerical investigations, owing to the exponential
growth of their Hilbert-space dimension with the number of degrees of
freedom. In recent years, substantial progress has been achieved through the introduction of new classes of many-body models and the development of powerful mathematical tools for their analysis. In the present work, we focus on dual-unitary quantum systems, which possess the remarkable property of being unitary with respect to both temporal and spatial evolution.

A prototypical model exhibiting this property, a chain of linearly
coupled Arnold cat maps, was introduced in Ref.~\cite{GutOsi15} and
subsequently investigated at both the classical and quantum levels
\cite{GHJSC16, LiangCvit20022, Fouxon_Gutkin_2022, GO2026}. Other
examples of dual-unitary systems have been identified in a variety of
settings, including kicked Ising spin chains and their generalizations
\cite{AWGG16, BeKoPr18, BeKoPr19-1, GBAWG20}, as well as quantum
circuit models \cite{BeKoPr19-4}. Although a complete characterization
of dual-unitary systems is still lacking, the existing constructions
already constitute a broad class of models that can be systematically
generated \cite{Arul19}.

Dual-unitary models have attracted considerable attention in the study of many-body quantum chaos~\cite{BeKoPr19-4, AWGG16, GopLam19, BeKoPr19-1, LakshPal2018, BeKoPr18, BWAGG19, BeKoPrPi19, BeKoPr2019operator, KPBBPT_2021, zhou2019entanglement, AVAN2016, Karl15, Arul19, Arul2021} due to remarkable combination of chaotic behavior and their analytical tractability. On the one hand, they display hallmarks of maximally chaotic many-body systems, including Wigner--Dyson spectral statistics and robustness against many-body localization effects \cite{BeKoPr18, BWAGG19, PhysRevResearch.3.023118}. On the other hand, dual-unitary models admit a wealth of exact analytical results; see \cite{BertinyClaeysProsen2026} for a recent review.
In particular, the interplay of duality and causality enables the exact computation of local two-point correlation functions~\cite{BeKoPr19-4, CL20, GBAWG20}. The analysis of correlation functions remains an active area of research~\cite{borsi2022construction, ippoliti2021postselection, claeys2022exact, KPBBPT_2021, krajnik2020kardar, lu2021spacetime, prosen2021many}. Other important directions include the investigation of matrix-product-state structures~\cite{piroli2020exact, lerose2021influence}, nonequilibrium steady states and eigenstate thermalization~\cite{ippoliti2022fractal, fritzsch2021eigenstate}, computational complexity and simulation methods~\cite{suzuki2022computational}, as well as random-matrix spectral statistics~\cite{flack2020statistics, bertini2021random}. More recently, dual-unitary dynamics have been realized on a quantum computing platform, demonstrating that present-day quantum processors can accurately simulate the many-body correlation functions~\cite{fisher2026Nature}.

Previous studies have focused predominantly on dual-unitary models
with nearest-neighbor interactions in one or more spatial dimensions
\cite{jonay2021triunitary, GutOsipovGuhr2024}. In such systems,
information propagates with unit velocity, and correlations between
local operators vanish identically everywhere except along the edges of the light cone, $|n| = |t|$. 

In the present work, we extend this framework to dual-unitary kicked
spin chains with an arbitrary interaction range $r$. At first sight,
the requirement of dual unitarity might appear to be highly
restrictive in the presence of long-range interactions. Remarkably, a kicked spin chain with general interactions involving
$r$ neighboring spins can be rendered dual-unitary by introducing an additional pairwise interaction between spins separated by $r-1$ sites. Since the number of possible interaction terms increases rapidly with $r$, this construction generates a broad and
highly flexible family of dual-unitary spin chains with long-range
interactions.

For kicked models with interactions extending over $r$ lattice sites,
the maximal velocity of information propagation is determined by the
interaction range. Consequently, local correlations propagate along
the light cone edges $|n|=r|t|$.
As we demonstrate in the main text, the correlation functions along
these lines  can be evaluated analytically. We
illustrate the general construction by considering a kicked Ising
chain (KIC) with next-to-next-nearest-neighbor interactions and derive
explicit expressions for its local two-point correlation functions.
  
\paragraph*{The paper is organized as follows:}
In the next section (Sec.~\ref{SecMainI}), we show how a generic class of kicked Ising chains
with long-range interactions can be constructed as a one-dimensional
representation of a multidimensional generalization of the standard
KIC. In Sec.~\ref{Sec3}, we introduce a broader class of such models and, in Sec.~\ref{LongRangeKICDualitySec}, prove their
dual-unitarity based on the Hadamard property of the local evolution matrices. In Sec.~\ref{CorrelationLongRangeKIC} we derive closed-form analytical expressions for connected correlation functions between local operators. In Sec.~\ref{Sec6} we apply this result for explicit calculation of the correlation function in the kicked spin chains with next-to-next-nearest-neighbor interactions. The explicit expressions for the correlation functions are gathered in Appendices~\ref{AppBoundNonHom}, and~\ref{AppBoundHom}. Finally, Sec.~\ref{Sec7} summarizes our main results and presents
our conclusions.

\section{Main ideas}\label{SecMainI}
The kicked Ising chain (KIC) is a paradigmatic model of periodically
driven many-body quantum system. It  belongs to the broader class of periodically
kicked quantum systems that originated in studies of the quantum
kicked rotor~\cite{GChIF1979} and the kicked top~\cite{haakeKT}. Originally, introduced  in the context of
exactly solvable quantum spin chains~\cite{Prosen1998}, it can be viewed as a Floquet
extension of the one-dimensional Ising model, in which the Ising
interaction is combined with periodic kicks generated by a transverse
magnetic field.  

A general model with stroboscopic evolution is governed by the
time-dependent Hamiltonian
\begin{equation}\label{Ham}
H(t)=H_I+H_K\sum_{\tau=-\infty}^{+\infty}\delta(t-\tau),
\end{equation}
where the kick Hamiltonian $H_K$ acts at integer times and generates
the evolution of the otherwise non-interacting spins, while the
interaction Hamiltonian $H_I$ governs the dynamics between consecutive
kicks. The stroboscopic time evolution over one driving period is
therefore described by the Floquet operator
\begin{equation}\label{Unitgeneric}
U=\hat{U}_{K}\hat{U}_{I},
\qquad
\hat{U}_{K}=e^{-iH_K},
\quad
\hat{U}_{I}=e^{-iH_I},
\end{equation}
where $\hat{U}_{K}$ and $\hat{U}_{I}$ are the unitary evolution
operators generated by the kick and interaction Hamiltonians,
respectively~\cite{AWGG16, AWGBG16, AGBWG18}.

\begin{figure}
\begin{center}
\begin{tabular}{lc}
    a)& \begin{tikzpicture}
             \def\xstep{1.5}
             \def\ystep{1.5}
             \def\arrowlen{0.6}

             \foreach \i in {0,1,2} {
                 \pgfmathsetmacro{\x}{\i*\xstep}
                 \pgfmathsetmacro{\xnext}{(\i+1)*\xstep}
                 \draw[red, very thick] (\x-0.6,\ystep) -- (\xnext+0.6,\ystep)
                 node[black, midway, above] {\small $J$};
                 \draw[red, very thick] (\x-0.6,0) -- (\xnext+0.6,0)
                 node[black, midway, below] {\small $J$};
             }

             \foreach \i in {0,1,2,3} {
                 \pgfmathsetmacro{\x}{\i*\xstep}

                 \draw[blue, very thick] (\x,0) -- (\x,\ystep)
                 node[black, midway, left] {\small $J_2$};

                 \filldraw[color=black, fill=white, very thick]
                 (\x,0) circle (8pt);
                 \node at (\x,0) {\small
                 \pgfmathparse{int(2*\i)}\pgfmathresult};

                 \filldraw[color=black, fill=white, very thick]
                 (\x,\ystep) circle (8pt);
                 \node at (\x,\ystep) {\small
                 \pgfmathparse{int(2*\i+1)}\pgfmathresult};
             }

\end{tikzpicture}
\\
   b)&\begin{tikzpicture}
     \def\xstep{1.5}
     \def\arrowlen{0.4}
     \def\ypos{0}
     \def\arcHeight{0.8}

     \foreach \i/\j in {0/2,1/3,2/4,3/5,4/6,5/7} {
         \pgfmathsetmacro{\xa}{\i*\xstep}
         \pgfmathsetmacro{\xb}{\j*\xstep}
         \draw[red,bend left=60, very thick] (\xa,\ypos) to (\xb,\ypos);
     }

     \pgfmathsetmacro{\xj}{(3*\xstep+4*\xstep)/2}
     \node[black] at (\xj,\ypos+1.1*\arcHeight) {\small $J$};

     \foreach \i/\j in {6/8,7/9} {
         \pgfmathsetmacro{\xa}{\i*\xstep}
         \pgfmathsetmacro{\xb}{\j*\xstep}
         \pgfmathsetmacro{\radius}{(\xb-\xa)/4}
         \draw[red, very thick]
         (\xa,\ypos) arc[start angle=180,end angle=90,radius=\radius];
     }

     \foreach \i/\j in {0/2,1/3} {
         \pgfmathsetmacro{\xa}{\i*\xstep}
         \pgfmathsetmacro{\xb}{\j*\xstep}
         \pgfmathsetmacro{\radius}{(\xb-\xa)/4}
         \draw[red, very thick]
         (\xa,\ypos) arc[start angle=0,end angle=90,radius=\radius];
     }

     \foreach \i/\j in {0/1,2/3,4/5,6/7} {
         \pgfmathsetmacro{\xa}{\i*\xstep}
         \pgfmathsetmacro{\xb}{\j*\xstep}
         \draw[blue, very thick] (\xa,\ypos) -- (\xb,\ypos);
         \node[black] at ({(\xa+\xb)/2},\ypos-0.3) {\small $J_2$};
     }

     \foreach \i in {0,...,7} {
         \pgfmathsetmacro{\x}{\i*\xstep}
         \filldraw[color=black, fill=white, very thick]
         (\x,\ypos) circle (8pt);
         \node at (\x,\ypos) {\small \i};
     }

\end{tikzpicture}
\\
   c)&\begin{tikzpicture}
     \def\xstep{1.5}
     \def\arrowlen{0.4}
     \def\ypos{0}
     \def\arcHeight{0.8}

     \foreach \i/\j in {0/2,1/3,2/4,3/5,4/6,5/7} {
         \pgfmathsetmacro{\xa}{\i*\xstep}
         \pgfmathsetmacro{\xb}{\j*\xstep}
         \draw[red,bend left=60, very thick] (\xa,\ypos) to (\xb,\ypos);
     }

     \pgfmathsetmacro{\xj}{(3*\xstep+4*\xstep)/2}
     \node[black] at (\xj,\ypos+1.1*\arcHeight) {\small $J$};

     \foreach \i/\j in {6/8,7/9} {
         \pgfmathsetmacro{\xa}{\i*\xstep}
         \pgfmathsetmacro{\xb}{\j*\xstep}
         \pgfmathsetmacro{\radius}{(\xb-\xa)/4}
         \draw[red, very thick]
         (\xa,\ypos) arc[start angle=180,end angle=90,radius=\radius];
     }

     \foreach \i/\j in {0/2,1/3} {
         \pgfmathsetmacro{\xa}{\i*\xstep}
         \pgfmathsetmacro{\xb}{\j*\xstep}
         \pgfmathsetmacro{\radius}{(\xb-\xa)/4}
         \draw[red, very thick]
         (\xa,\ypos) arc[start angle=0,end angle=90,radius=\radius];
     }

     \foreach \i/\j in {0/1,2/3,4/5,6/7} {
         \pgfmathsetmacro{\xa}{\i*\xstep}
         \pgfmathsetmacro{\xb}{\j*\xstep}
         \draw[blue, very thick] (\xa,\ypos) -- (\xb,\ypos);
         \node[black] at ({(\xa+\xb)/2},\ypos-0.3) {\small $J_2$};
     }

     \foreach \i/\j in {1/2,3/4,5/6} {
         \pgfmathsetmacro{\xa}{\i*\xstep}
         \pgfmathsetmacro{\xb}{\j*\xstep}
         \draw[yellow, very thick] (\xa,\ypos) -- (\xb,\ypos);
         \node[black] at ({(\xa+\xb)/2},\ypos-0.3) {\small $J_1$};
     }

     \foreach \i in {0,...,7} {
         \pgfmathsetmacro{\x}{\i*\xstep}
         \filldraw[color=black, fill=white, very thick]
         (\x,\ypos) circle (8pt);
         \node at (\x,\ypos) {\small \i};
     }

\end{tikzpicture}
\end{tabular}
\end{center}
\caption{\small
(a) A single layer of the two-dimensional KIC with nearest-neighbor
interactions. The interactions are shown by colored lines connecting the particles depicted by circles with numbers. The interaction strength is shown by the corresponding symbols $J$, or $J_2$. (b) The equivalent one-dimensional spin chain with
next-nearest-neighbor interactions corresponding to $r=2$, see the Hamiltonian in eqs.~(\ref{HI1},~\ref{mmodel1}) taken at $r=2$. The model
is dual-unitary for $J=\pi/4$; (c) A generalization of the
constructions in (a) and (b): a spin chain with next-nearest-neighbor
interactions corresponding to $r=2$ and two types of nearest-neighbor
interactions' strengths $J_1$ and $J_2$.
}
\label{Bild12}
\end{figure}

The standard KIC  describes a chain of $N$
spin-$1/2$ degrees of freedom with nearest-neighbor Ising interactions,
\begin{equation}
    H_{I}
    =
    J \sum_{n=1}^{N}
    \hat{\sigma}_{n}^{z}\hat{\sigma}_{n+1}^{z}
    +
    d \sum_{n=1}^{N}\hat{\sigma}_{n}^{z},
    \label{interactionKIC}
\end{equation}
subject to periodic kicks generated by a transverse magnetic field,
\begin{equation}
    H_{K}
    =
    h_x \sum_{n=1}^{N}
    \hat{\sigma}_{n}^{x},
    \label{kickKIC}
\end{equation}
where  the local spin operators
\[ \hat\sigma_n^\alpha=\underbrace{\mathds{1}\otimes  \dots\otimes \mathds{1}}_{n -1} \otimes\,  \sigma^\alpha\otimes \underbrace{\mathds{1}\otimes  \dots\otimes \mathds{1}}_{N-n },\]
are defined in terms of the Pauli matrices $\sigma^\alpha$, $\alpha=x,y,z$, with periodic boundary conditions imposed by identifying $\hat{\sigma}_{N+1}^\alpha=\hat{\sigma}_{1}^\alpha$.
The model becomes dual-unitary at the self-dual point~\cite{AWGG16},
\begin{equation}\label{sdpoint}
    h_x=J=\frac{\pi}{4}.
\end{equation}
Remarkably, for this choice of parameters, the kicked Ising chain becomes, in a sense, a solvable model of quantum chaos. In particular, it allows for the exact evaluation of the short-time spectral form factor~\cite{BeKoPr18}, as well as the exact calculation of correlations between local operators~\cite{BeKoPr19-4, CL20, GBAWG20}.

A natural question is whether the interaction range of the KIC can be
extended while preserving dual-unitarity. To investigate this
possibility, we first consider a two-dimensional extension of the KIC,
introduced and studied in Ref.~\cite{GutOsipovGuhr2024}. The
corresponding Hamiltonians describe a two-dimensional lattice of spins
with nearest-neighbor interactions,
\begin{equation}\label{HI2D}
    H_{I}
    =
    \sum_{n,m}
    J_{1}(n,m)
    \hat{\sigma}_{n,m}^{z}\hat{\sigma}_{n+1,m}^{z}
    +
    J_{2}(n,m)
    \hat{\sigma}_{n,m}^{z}\hat{\sigma}_{n,m+1}^{z}
    +
    d(n,m)\hat{\sigma}_{n,m}^{z},
\end{equation}
subject to on-site kicks,
\begin{equation}\label{HK2D}
    H_{K}
    =
    h_x\sum_{n,m}\hat{\sigma}_{n,m}^{x}.
\end{equation}

As shown in Ref.~\cite{GutOsipovGuhr2024}, for
$J_{1}(n,m)=h_x=\pi/4$ and arbitrary values of $J_{2}(n,m)$, the model
is partially dual-unitary. In particular, this implies that the
correlation function
\begin{equation}
   C(n,m,t)=
   \frac{1}{2^{NM}}
   \trace\!\left(
   U^t Q_1 U^{-t} Q_2
   \right)
\end{equation}
between two local operators $Q_1$ and $Q_2$, supported at the lattice
sites $(0,0)$ and $(n,m)$, respectively, can be nonzero only along the
lines
\[
|n|=|t|,
\qquad
m=0.
\]
Moreover, these correlations can be evaluated analytically when the
operators are supported on a plaquette of four neighboring lattice
sites. Further analysis revealed that, for
$
J_{2}(n,m)=\frac{\pi}{4},
$
the model possesses complete space--time duality. Consequently, all
local two-point correlation functions vanish identically for times
$t$ exceeding the spatial extent of the supports of the operators
$Q_1$ and $Q_2$.

The two-dimensional lattice model defined by
Eqs.~(\ref{HI2D}) and~(\ref{HK2D}) provides a convenient route to
constructing a long-range dual-unitary extension of the KIC; see
Fig.~\ref{Bild12}. To this end, we exploit the freedom in the choice
of $J_{2}(n,m)$ and set
\begin{equation}
    J_2(n,m)=
\left\{
\begin{array}{ll}
J_2, & m=0,1,\ldots,r-1,\\
0, & \mathrm{otherwise}.
\end{array}
\right.
\end{equation}
This choice defines a strip of width $r$ that is decoupled from the
remaining lattice sites. By enumerating the spins within the strip
according to
$
k=rn+m,
$
the strip can be mapped onto a one-dimensional spin chain of length
$
\mathcal{N}=rN.
$
The interaction Hamiltonian then consists of two contributions,
$H_I=H_I^{(1)}+H_I^{(2)}$, where
\begin{equation}\label{HI1}
    H_{I}^{(1)}
    =
    \frac{\pi}{4} \sum_{n}
    \hat{\sigma}_{n}^{z}\hat{\sigma}_{n+r}^{z}
    ,
\end{equation}
and
\begin{equation} 
    H_{I}^{(2)}
    =
    J_{2}
    \sum_{n\bmod r\neq 0}
    \hat{\sigma}_{n}^{z}\hat{\sigma}_{n+1}^{z}+
    d\sum_{n}\hat{\sigma}_{n}^{z},
    \label{model1}
\end{equation}
together with the kick Hamiltonian~(\ref{kickKIC}). The case $r=2$ is
illustrated in Fig.~\ref{Bild12}b.

By construction, the resulting kicked spin chain is dual-unitary for arbitrary values of $J_2$ and $d$. The parameter $r$ determines the
interaction range and, correspondingly, the maximal velocity of
information propagation in the spin chain. The standard KIC is
recovered as the special case $r=1$.

\begin{figure}
\includegraphics[width=1.0\linewidth]{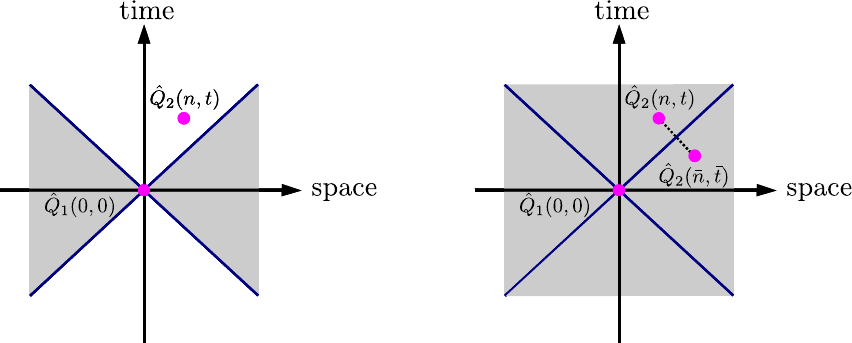}
\caption{\small
Illustration of the correlations
$\langle \hat{Q}_1(0,0)\hat{Q}_2(n,t)\rangle$ for two traceless observables,  $\langle \hat{Q}_1\rangle=\langle \hat{Q}_2\rangle=0$, localized at the space-time points $(0,0)$ and $(n,t)$, respectively. For a general kicked model with interactions extending over $r$  neighboring lattice sites, the maximal information propagation velocity is $r$. Therefore, correlations in the shaded regions, $|n|>r|t|$, vanish by causality, as shown in the left panel. For a dual-unitary model, defined by the Hamiltonians (\ref{HI1}, \ref{model1}) at the choice~(\ref{sdpoint}), one has in addition 
$\langle \hat{Q}_1(0,0)\hat{Q}_2(n,t)\rangle
=\langle \hat{Q}_1(0,0)\hat{Q}_2(\bar{n},\bar{t})\rangle$,
where $\bar{n}=rt$ and $\bar{t}=n/r$.
Consequently, by causality, the correlations also vanish at the points satisfying $|\bar{n}|>r|\bar{t}|$, or equivalently, in the interior of the light cone, $|n|<r|t|$, as shown in the right panel. Thus, nonzero correlations can occur only along the light-cone edges, $|n|=r|t|$, indicated by the blue lines. 
}
\label{Bild1}
\end{figure}

The correspondence between the two-dimensional spin lattice and the
associated one-dimensional spin-chain representation leads to a simple
prediction for the propagation of local correlations. Let $Q_1$ and
$Q_2$ be local traceless operators supported at sites $0$ and
$
k=rn+m
$
of the spin chain, respectively. In the original two-dimensional
lattice, these sites correspond to the coordinates $(0,0)$ and
$(n,m)$. According to the results of Ref.~\cite{GutOsipovGuhr2024},
correlations between these operators are nonzero only along the lines
$
|n|=|t|$,
$m=0$.
Therefore, in the spin-chain representation, local correlations
propagate exclusively along the edges of the light cone,
$
|k|=r|t|,
$
as illustrated in Fig.~\ref{Bild1}.

\section{A family of long-range  dual-unitary kicked chain}\label{Sec3}

In contrast to the standard KIC, the interactions in the spin-chain Hamiltonian~(\ref{HI1},~\ref{model1}) are not homogeneous. Instead, they are invariant under translations by $r$ lattice sites. As we show below, this model represents a particular member of a broader family of long-range kicked Ising chains, which includes both homogeneous and inhomogeneous dual-unitary models.

\subsection{Dual-unitary long-range kicked spin chain}

A generalization of the model introduced in the previous section, cf.~eqs.~(\ref{HI1}) and~(\ref{model1}), can be obtained by replacing the nearest-neighbor interaction term with  an arbitrary function that couples $r$ consecutive  spins. More precisely, we consider
\begin{equation}
    H_{I}^{(2)}
    =
    \sum_{n=1}^{N}
    f_n\left(
        \hat{\sigma}_{n}^{z},
        \hat{\sigma}_{n+1}^{z},
        \ldots,
        \hat{\sigma}_{n+r-1}^{z}
    \right),
    \label{mmodel1}
\end{equation}
where $f_n$ are, in general, position-dependent functions of $r$ spin variables. Periodic boundary conditions are imposed according to
$
    \hat{\sigma}_{n+N}^{z}=\hat{\sigma}_{n}^{z}.
$
The kick Hamiltonian is given by Eq.~(\ref{kickKIC}).

As we demonstrate in Sec.~\ref{LongRangeKICDualitySec}, the resulting model is dual-unitary for $H_{I}^{(1)}$ given by Eq.~(\ref{HI1}), and $H_{I}^{(2)} $
with  an arbitrary choice of the functions $f_n$ in Eq.~(\ref{mmodel1}). Thus, dual unitarity does  allows for a broad class of position-dependent couplings in the presence of long-range interactions.

For $r=2$, a particular realization of this family is obtained by choosing (see Fig.~\ref{Bild12}c)
\begin{equation}
f_n\left(
    \hat{\sigma}_{n}^{z},
    \hat{\sigma}_{n+1}^{z}
\right)
=
\left\{
\begin{array}{ll}
J_2\,\hat{\sigma}_{n}^{z}\hat{\sigma}_{n+1}^{z}+d\hat{\sigma}_{n}^{z},
& n\ \mathrm{even},\\[2mm]
J_1\,\hat{\sigma}_{n}^{z}\hat{\sigma}_{n+1}^{z}+d\hat{\sigma}_{n}^{z},
& n\ \mathrm{odd}.
\end{array}
\right.
\end{equation}
The Hamiltonian~(\ref{model1}) is recovered as the special case
$
    J_1=0.
$
Another important special  case, which will be considered in detail below, is obtained for
$
    J_1=J_2,
$
corresponding to a homogeneous spin chain with long-range interactions.

\subsection{Dual-unitary long-range kicked chain with arbitrary on-site Hilbert-space dimension}

The standard KIC has the smallest nontrivial on-site Hilbert-space dimension, $L=2$. Here, we generalize the construction to an arbitrary on-site dimension $L$. This extension is particularly useful because it allows one to construct families of models that admit a well-defined classical limit.

Our construction generalizes the one introduced in Ref.~\cite{GBAWG20} and is based on a pair of $L\times L$ complex Hadamard matrices,
\begin{equation}\label{Hadamard}
    \bra{s}u[h]\ket{\bar{s}}
    =\frac{e^{i h(s,\bar{s})}}{\sqrt{L}},
    \qquad
    \bra{s}u[g]\ket{\bar{s}}
    =\frac{e^{i g(s,\bar{s})}}{\sqrt{L}},
\end{equation}
where $h(s,\bar{s})$ and $g(s,\bar{s})$ are real-valued functions. By construction, all matrix elements of $u[f]$ and $u[g]$ have the same absolute value, $L^{-1/2}$, while the matrices themselves are unitary.

To specify the Floquet evolution, we first introduce the on-site Hilbert space $\mathcal{H}$ with the discrete $L$-dimensional basis
\begin{equation}
    \left\{
    \ket{s},\; s=\overline{1,L}
    \right\}.
\end{equation}
The Hilbert space of the $N$-site chain is then given by
$
    \mathcal{H}_{N}=\mathcal{H}^{\otimes N},
$
and has dimension $L^N$. Its natural product basis is
\begin{equation}\label{basis}
    \left\{
    \ket{\bm{s}}
    \equiv
    \bigotimes_{j=1}^{N}\ket{s_j},
    \qquad
    s_j=\overline{1,L}
    \right\},
\end{equation}
where the index $j$ labels the sites of the chain.

Given the pair of complex Hadamard matrices in eq.~(\ref{Hadamard}), we define the Floquet evolution as a product of a local kick and an interaction evolution. The kick operator is the tensor product
\begin{equation}\label{UKgdef}
    U_K[g]=\bigotimes_{n=1}^{N}u[g],
\end{equation}
whose matrix elements in the product basis~(\ref{basis}) are
\begin{equation}
    \braket{\bm{s}|U_K[g]|\bm{s}'}
    =
    \left(\frac{1}{L}\right)^{N/2}
    \exp\left[
        i\sum_{n=1}^{N}g(s_n,s_n')
    \right].
\end{equation}
The interaction operator is diagonal in the product basis and is defined by
\begin{equation}\label{UIFloquet}
    \braket{\bm{s}|U_I[h,\bm{f}]|\bm{s}'}
    =
    \delta_{\bm{s},\bm{s}'}
    \exp\left[
        i\sum_{n=1}^{N}
        \left(
        h(s_n,s_{n+r})
        +
        f_n(s_n,s_{n+1},\ldots,s_{n+r-1})
        \right)
    \right].
\end{equation}
Here, the functions $h$ and $g$ are constrained by the complex-Hadamard condition~(\ref{Hadamard}), whereas the set of functions
\begin{equation}\label{fbold}
    \bm{f}=(f_1,f_2,\ldots,f_N)
\end{equation}
is otherwise arbitrary. Thus, the functions $f_n$ provide a considerable freedom in constructing inhomogeneous long-range interactions without affecting the dual-unitary property.

For the standard KIC ($L=2$), the  functions $g$, $h$ and  $f$ are given by
\begin{equation}\label{ghdualunitarychoice}
    g(s,\bar{s})=-h(s,\bar{s})=
    \frac{\pi}{4}
    \left[
        (2s-3)(2\bar{s}-3)-1
    \right], \quad 
    f(s)=-d(2s-3)-\pi/4,
\end{equation}
where the arguments $s$, and $\bar{s}$ take the values $1$ and $2$.  In this case  $u[g], u[h]$, are  $2\times2$ complex Hadamard matrices and, up to a conjugation by diagonal matrices, coincide with the two-dimensional discrete Fourier transform. For arbitrary $L$, the discrete Fourier transform,
\begin{equation}
    \bra{s}u\ket{\bar{s}}
    =
    \frac{1}{\sqrt{L}}
    \exp\left(
        \frac{2\pi i}{L}(s-1)(\bar{s}-1)
    \right),
\end{equation}
provides a canonical example of a complex Hadamard matrix and therefore gives a natural extension of the construction to higher-dimensional on-site Hilbert spaces.

As observed in Ref.~\cite{GBAWG20}, for $r=1$ the resulting model can be interpreted as a quantization of a chain of linearly coupled Arnold cat maps, whose classical dynamics has been studied extensively~\cite{GutOsi15,LiangCvit20022,Fouxon_Gutkin_2022}. For $r\geq2$, the present construction yields a new class of coupled cat-map models with long-range interactions that retain the dual-unitary property. Investigating the corresponding classical dynamics and its implications for the quantum properties of these systems would be of considerable interest, but lies beyond the scope of the present work.

\section{Spatio-temporal duality of the long-range KIC models}
\label{LongRangeKICDualitySec}

The unitary operator defined by Eqs.~(\ref{UKgdef})--(\ref{fbold}),
\begin{equation}\label{Ugeneric}
    U=\hat{U}_{K}[g]\hat{U}_{I}[h,\bm f],
\end{equation}
describes the temporal evolution of a system with Hilbert-space dimension
$L^N$, where $N$ is the number of particles. In complete analogy with the temporal evolution, one can introduce a dual spatial evolution operator. To this end, we consider the dual Hilbert space
$\mathcal{H}^{\otimes rT}$ of dimension $L^{rT}$, where $T$ denotes the total number of time steps. We equip this space with the product basis
\begin{equation}
    \label{basisdual}
    \ket{\bm{\eta}}
    \equiv
    \bigotimes_{t=1}^{T}\ket{\bm{\eta}_t},
    \qquad
    \ket{\bm{\eta}_t}
    \equiv
    \bigotimes_{j=1}^{r}\ket{\eta_j^t},
    \qquad
    \eta_j^t=\overline{1,L}
.
\end{equation}

The corresponding dual evolution operator is defined as
\begin{equation}
    \tilde{U}
    =
    \tilde{U}_{K}[h,\bm f]\tilde{U}_{I}[g],
\end{equation}
and has a structure analogous to that of the original Floquet operator, but acts on the dual Hilbert space of dimension $L^{rT}$.

The interaction part of the dual evolution operator is diagonal in the product basis,
\begin{equation}
    \label{UIFloquet2}
    \braket{\bm{\eta}|\tilde{U}_{I}[g]|\bm{\eta}'}
    =
    \delta(\bm{\eta},\bm{\eta}')
    e^{
        i\sum_{t=1}^{T}\sum_{j=1}^{r}
        g(\eta_j^t,\eta_j^{t+1})
    },
\end{equation}
where, as before, periodic boundary conditions in the temporal direction are imposed,
\begin{equation}
    \eta_j^{T+t}\equiv\eta_j^t.
\end{equation}
The dual kick operator is a tensor product of identical local operators,
\begin{equation}
    \tilde{U}_{K}[h,\bm f_n]
    =
    \bigotimes_{t=1}^{T}
    \tilde{u}[h,\bm f_n],
\end{equation}
where the local operator $\tilde{u}[h,\bm f_n]$ acts on the $L^r$-dimensional local Hilbert space according to
\begin{equation}\label{dualop}
    \braket{\bm{\eta}_t|
    \tilde{u}[h,\bm f_n]|
    \bar{\bm{\eta}}_t}
    =
    \left(\frac{1}{L}\right)^{r/2}
    e^{
    i\sum_{j=1}^{r}
    \Big(
    h(\eta_j^t,\bar{\eta}_j^t)
    +f_{n+j}
    \big(
    \eta_j^t,\ldots,\eta_r^t,
    \bar{\eta}_1^t,\ldots,\bar{\eta}_{j-1}^t
    \big)
    \Big)
    }.
\end{equation}
Here,
$
    \bm f_n=(f_{n+1},\ldots,f_{n+r})
$
denotes the corresponding set of $r$ functions selected from the collection
$\bm f$. The  dependence on $n$ reflects the possible spatial inhomogeneity of the original long-range KIC.

The construction above possesses a dual-unitarity property analogous to that of the standard KIC. This result follows from the two propositions below.

\medskip

\noindent\textbf{Proposition 1.}
For arbitrary propagation time $T$ and a chain length $N$ divisible by $r$, the traces of the temporal and dual spatial evolution operators satisfy
\begin{equation}
    \tr U^T
    =
    \tr
    \left(
    \prod_{n=1}^{N/r}\tilde{U}_n
    \right).
    \label{dualitytrrelation}
\end{equation}
In particular, for the homogeneous case, in which the functions $f_n$ are independent of $n$ and hence
$\tilde{U}_n=\tilde{U}$, this relation reduces to
\begin{equation}
    \label{duality}
    \tr U^T
    =
    \tr\tilde{U}^{N/r}.
\end{equation}

\medskip

\noindent\textit{Proof.}
The proof of Proposition~1 is given in Appendix~\ref{AppA}.
\hfill\ensuremath{\Box}

\medskip

The unitarity of $U$ follows directly from its construction. It turns out that the corresponding dual operators $\tilde{U}_n$ are also unitary. This is the key property underlying the  space--time duality of the model.

\medskip

\noindent\textbf{Proposition 2.}
For an arbitrary set of functions $\bm f$ and arbitrary functions $h$ and $g$ associated with a pair of complex Hadamard matrices as defined in Eq.~(\ref{Hadamard}), the dual operator $\tilde{U}_n$ is unitary.

\medskip

\noindent\textit{Proof.}
Since the dual evolution operator is a tensor product of local dual kicks and diagonal unitary interaction operators, it is sufficient to prove the unitarity of the local operator $\tilde{u}[h,\bm f_n]$. We therefore consider
\begin{equation}
    \sum_{\bm{\eta}}
    \braket{\bm{s}|\tilde{u}[h,\bm f_n]|\bm{\eta}}
    \braket{\bm{\eta}|\tilde{u}[h,\bm f_n]|\bm{s}'}
    =
    \delta(\bm{s},\bm{s}').
    \label{unitarity}
\end{equation}
Using Eq.~(\ref{dualop}), the left-hand side becomes
\begin{equation}\label{unitarity1}
\left(\frac{1}{L}\right)^{r}\sum_{\eta_1,\dots,\eta_r}
e^{i\sum_{t=1}^r
\Big(h(s_t,\eta_t)-h(s'_t,\eta_t)
+f_{n+t}(s_t,\cdots,s_r,\eta_1,\cdots,\eta_{t-1})
-f_{n+t}(s'_t,\cdots,s'_r,\eta_1,\cdots,\eta_{t-1})\Big)}.
\end{equation}

The crucial observation is that the variable $\eta_r$ appears in the exponent only through the combination
$    h(s_r,\eta_r)-h(s'_r,\eta_r) $.
Consequently, the sum over $\eta_r$ can be evaluated using the unitarity of the complex Hadamard matrix $u[h]$,
\begin{equation}
    \frac{1}{L}
    \sum_{\eta_r}
    e^{i[h(s_r,\eta_r)-h(s'_r,\eta_r)]}
    =
    \delta(s_r,s'_r).
\end{equation}
Equation~(\ref{unitarity1}) therefore reduces to
\begin{multline}\label{unitarity2}
\left(\frac{1}{L}\right)^{r-1}
    \sum_{\eta_1,\ldots,\eta_{r-1}}
    e^{
    i\sum_{t=1}^{r-1}
    \Big(
    h(s_t,\eta_t)-h(s'_t,\eta_t)
    +f_{n+t}(s_t,\ldots,s_{r-1},\eta_1,\ldots,\eta_{t-1})
    -f_{n+t}(s'_t,\ldots,s'_{r-1},\eta_1,\ldots,\eta_{t-1})
    \Big)
    }\\\times
    \delta(s_r,s'_r).
\end{multline}

The same argument can then be applied successively to the remaining variables. At the next step, the sum over $\eta_{r-1}$ yields
\begin{multline}\label{unitarity3}
\left(\frac{1}{L}\right)^{r-2}
    \sum_{\eta_1,\ldots,\eta_{r-2}}
    e^{
    i\sum_{t=1}^{r-2}
    \Big(
    h(s_t,\eta_t)-h(s'_t,\eta_t)
    +f_{n+t}(s_t,\ldots,s_{r-2},\eta_1,\ldots,\eta_{t-1})
    -f_{n+t}(s'_t,\ldots,s'_{r-2},\eta_1,\ldots,\eta_{t-1})
    \Big)
    }\\\times
    \delta(s_{r-1},s'_{r-1})
    \delta(s_r,s'_r).    
\end{multline}

Repeating this procedure successively for
$\eta_r,\eta_{r-1},\ldots,\eta_1$
eliminates all summation variables and produces
\begin{equation}
    \delta(s_1,s'_1)\cdots\delta(s_r,s'_r)
    =
    \delta(\bm{s},\bm{s}'),
\end{equation}
which proves the unitarity of
$\tilde{u}[h,\bm f_n]$ and, consequently, of $\tilde{U}_n$.
\hfill\ensuremath{\Box}

\medskip

\noindent\textbf{Remark.}
The local dual kick admits the factorized representation
\begin{equation}
    \tilde{u}[h,\bm f_n]
    =
    \left(
    \bigotimes_{j=1}^{r}u[h]
    \right)
    D[\bm f_n],
    \label{decomp33}
\end{equation}
where $D[\bm f_n]$ is the diagonal unitary operator
\begin{equation}
    \label{dualop33}
    \braket{\bm{\eta}_t|
    D[\bm f_n]|
    \bar{\bm{\eta}}_t}
    =
    \left(
    \prod_{j=1}^{r}
    \delta(\eta_j^t,\bar{\eta}_j^t)
    \right)
    e^{
    i\sum_{j=1}^{r}
    f_{n+j}
    \big(
    \eta_j^t,\ldots,\eta_r^t,
    \bar{\eta}_1^t,\ldots,\bar{\eta}_{j-1}^t
    \big)
    }.
\end{equation}
Since the tensor product
$\bigotimes_{j=1}^{r}u[h]$
is a complex Hadamard matrix, multiplication by the diagonal unitary
$D[\bm f_n]$ preserves the modulus of every matrix element. Thus,
$\tilde{u}[h,\bm f_n]$ is itself a complex Hadamard matrix.

Propositions~1 and~2 establish the unitarity of both spatial and temporal evolutions. Importantly, the functions $f_n$ entering the construction are completely arbitrary. The resulting family therefore provides a broad class of dual-unitary kicked spin chains, including both homogeneous and inhomogeneous models with long-range interactions.
 
\section{Correlation functions of long-range KIC models}
\label{CorrelationLongRangeKIC}

The complete space--time duality of the model allows the local correlation functions properties are known, as well as their simplified form in terms of transfer operator, which allows in some practical cases evaluate them explicitly. We define the correlation function as the normalized trace of the product of the forward and backward time evolutions of two local observables, $\bar{\Sigma}$ and $\ubar{\Sigma}$,
\begin{equation}
    C(t)
    =
    L^{-N}\,
    \tr
    \left(
    \bar{\Sigma}\,
    U^{-t}\,
    \ubar{\Sigma}\,
    U^t
    \right).
    \label{CorrFunc}
\end{equation}
The observables $\bar{\Sigma}$ and $\ubar{\Sigma}$ are constructed as tensor products of the identity operator and two local operators, $\bm Q_1$ and $\bm Q_2$,
\begin{eqnarray}\label{sigmaQ1}
    \bar{\Sigma}
    &=&
    \bm Q_1\otimes
    \underbrace{\1\otimes\cdots\otimes\1}_{N-k},
     \\
    \ubar{\Sigma}
    &=&
    \underbrace{\1\otimes\cdots\otimes\1}_{n}
    \otimes \bm Q_2\otimes
    \underbrace{\1\otimes\cdots\otimes\1}_{N-n-k},
    \label{sigmaQ2}
\end{eqnarray}
where $\1$ denotes the identity operator on the single-site Hilbert space. The operators $\bm Q_1$ and $\bm Q_2$ are assumed to be traceless and to act nontrivially on $k$ consecutive sites; hence, each acts on a Hilbert space of dimension $L^k$. The integer $n$ specifies the spatial separation between the supports of the two operators.

For the dual-unitary models considered here, causality together with space--time duality imposes strong constraints on the correlation function~(\ref{CorrFunc}). In particular, for traceless local operators $\bm Q_1$ and $\bm Q_2$, the correlation function vanishes identically away from the edges of the light cone. Non-vanishing correlations can occur only when the operators are connected by a light-cone edge. For the long-range model with interaction range $r$, this condition reads
\begin{equation}
    |n|=rt.
    \label{lightconecondition}
\end{equation}
Thus, the correlations propagate ballistically with velocity $r$.

Along the light-cone edge, $n=\pm rt$, the correlation function can be expressed in terms of a transfer operator $\hat{\bm T}$ as
\begin{equation}
    C(t)
    =
    \bra{\bar{\bm\Phi}_{\bar{\Sigma}}}
    \hat{\bm T}^{\,t-2}
    \ket{\bm\Phi_{\ubar{\Sigma}}},
    \label{CD1}
\end{equation}
where the boundary vectors
$\bra{\bar{\bm\Phi}_{\bar{\Sigma}}}$ and
$\ket{\bm\Phi_{\ubar{\Sigma}}}$
are determined by the local operators $\bm Q_1$ and $\bm Q_2$, respectively.

For simplicity, in the remainder of the paper we restrict the explicit calculation to the case $r=2$. The generalization to larger interaction ranges follows along the same lines, with an appropriately enlarged transfer operator.

For generic local operators $\bar{\Sigma}$ and $\ubar{\Sigma}$ with four-site support, the corresponding $L^4\times L^4$ transfer matrix was derived in Ref.~\cite{GutOsipovGuhr2024}. In the product basis, its matrix elements are given by
\begin{equation}
    \bra{\chi,\eta,\chi_1,\eta_1}
    \hat{\bm T}
    \ket{\chi',\eta',\chi'_1,\eta'_1}=
    \frac{1}{L^6}
    \left|
    \sum_{s,\bar{s}=1}^{L}
    e^{i\mathcal{F}(s,\bar{s})}
    \right|^2 .
\label{That16x16}
\end{equation}
Here, the phase function $\mathcal{F}(s,\bar{s})$ is
\begin{eqnarray}
\mathcal{F}(s,\bar{s})
&=&
h(s,\chi)
+h(\eta',s)
+h(\bar{s},\chi_1)
+h(\eta'_1,\bar{s})
\nonumber\\
&&
+g(s,\chi')
+g(\eta,s)
+g(\bar{s},\chi'_1)
+g(\eta_1,\bar{s})
\nonumber\\
&&
+f_1(\bar{s},\chi)
+f_2(s,\bar{s})
+f_1(\eta'_1,s),
\label{Faction}
\end{eqnarray}
where $f_1$ and $f_2$ denote the functions $f_n$ associated with odd and even lattice sites, respectively; see Fig.~\ref{Fig1}.

The explicit form of the transfer matrix~(\ref{That16x16}) provides a convenient starting point for deriving analytical expressions for the correlation functions. In the following, we exploit its spectral decomposition to identify the contributions of the relevant eigenvalues and eigenvectors to the propagation of local correlations along the light-cone edges.

\begin{figure}
\begin{tikzpicture}
     \def\xstep{1.5}
     \def\arrowlen{0.4}
     \def\ypos{0}
     \def\yshift{0.7}
     \def\arcHeight{0.8}

        \foreach \i/\j in {1/3, 3/5} {
         \pgfmathsetmacro{\xa}{\i*\xstep}
         \pgfmathsetmacro{\xb}{\j*\xstep}
         \draw[red, very thick] (\xa,\ypos+\yshift) to (\xb,\ypos+\yshift);
     }

     \foreach \i/\j in {0/1, 2/3, 4/5} {
         \pgfmathsetmacro{\xa}{\i*\xstep}
         \pgfmathsetmacro{\xb}{\j*\xstep}
         \draw[blue, very thick] (\xa,\ypos) -- (\xb,\ypos+\yshift);
     }

     \foreach \i/\j in {1/2, 3/4} {
         \pgfmathsetmacro{\xa}{\i*\xstep}
         \pgfmathsetmacro{\xb}{\j*\xstep}
         \draw[yellow, very thick] (\xa,\ypos+\yshift) -- (\xb,\ypos);
     }
     
        \draw[green, very thick] (2*\xstep,\ypos) -- (2*\xstep,\ypos+\xstep);
        \draw[green, very thick] (3*\xstep,\ypos+\yshift) -- (3*\xstep,\ypos+\xstep+\yshift);
        \draw[green, very thick] (2*\xstep,\ypos) -- (2*\xstep,\ypos-\xstep);
        \draw[green, very thick] (3*\xstep,\ypos+\yshift) -- (3*\xstep,\ypos-\xstep+\yshift);

 \foreach \i/\j in {0/2, 2/4} {
         \pgfmathsetmacro{\xa}{\i*\xstep}
         \pgfmathsetmacro{\xb}{\j*\xstep}
         \draw[red, very thick] (\xa,\ypos) to (\xb,\ypos);
     }

        \filldraw[color=black, fill=white, very thick] (0,\ypos) circle (8pt);
        \filldraw[color=black, fill=white, very thick] (1*\xstep,\ypos+\yshift) circle (8pt);

        \filldraw[black] (2*\xstep,\ypos) circle (8pt);
        \filldraw[black] (3*\xstep,\ypos+\yshift) circle (8pt);

        \filldraw[color=black, fill=white, very thick] (4*\xstep,\ypos) circle (8pt);
        \filldraw[color=black, fill=white, very thick] (5*\xstep,\ypos+\yshift) circle (8pt);

        \filldraw[color=black, fill=white, very thick] (2*\xstep,\ypos+\xstep) circle (8pt);
        \filldraw[color=black, fill=white, very thick] (3*\xstep,\ypos+\xstep+\yshift) circle (8pt);

        \filldraw[color=black, fill=white, very thick] (2*\xstep,\ypos-\xstep) circle (8pt);
        \filldraw[color=black, fill=white, very thick] (3*\xstep,\ypos-\xstep+\yshift) circle (8pt);
         \node at (0*\xstep, \ypos) {\small $\eta'$};
         \node at (1*\xstep, \ypos+\yshift) {\small $\eta_1'$};

         \node at (2*\xstep, \ypos) {\small\color{white} $s_1$};
         \node at (3*\xstep, \ypos+\yshift) {\small\color{white} $s_2$};

         \node at (4*\xstep, \ypos) {\small $\chi$};
         \node at (5*\xstep, \ypos+\yshift) {\small $\chi_1$};

         \node at (2*\xstep, \ypos+\xstep) {\small $\eta$};
         \node at (3*\xstep, \ypos+\xstep+\yshift) {\small $\eta_1$};

         \node at (2*\xstep, \ypos-\xstep) {\small $\chi'$};
         \node at (3*\xstep, \ypos-\xstep+\yshift) {\small $\chi_1'$};

        \draw[red, very thick] (6*\xstep,\ypos+\xstep/3) -- (7*\xstep,\ypos+\xstep/3);
        \node at (7.5*\xstep, \ypos+\xstep/3) {\small $h(s,\bar{s})$};
        \draw[green, very thick] (6*\xstep,\ypos-0*\xstep/3) -- (7*\xstep,\ypos-0*\xstep/3);
        \node at (7.5*\xstep, \ypos-0*\xstep/3) {\small $g(s,\bar{s})$};
        \draw[yellow, very thick] (6*\xstep,\ypos-\xstep/3) -- (7*\xstep,\ypos-\xstep/3);
        \node at (7.5*\xstep, \ypos-\xstep/3) {\small $f_1(s,\bar{s})$};
        \draw[blue, very thick] (6*\xstep,\ypos-2*\xstep/3) -- (7*\xstep,\ypos-2*\xstep/3);
        \node at (7.5*\xstep, \ypos-2*\xstep/3) {\small $f_2(s,\bar{s})$};

\end{tikzpicture}
\caption{\small The scheme of indexes used to construct the transfer matrix, eqs.~(\ref{That16x16},~\ref{Faction}).}\label{Fig1}
\end{figure}

\section{Explicit results for $r=2$  KIC model}\label{Sec6}
In order to illustrate the above generic results we have chosen as an example the dual-unitary KIC with the next-to-next-neighbor ($r=2$, $L=2$) interactions formulated for two cases with non-homogeneous and homogeneous interactions. These two cases differ by the choice of the interaction Hamiltonians. In the dual-unitary regime at $r=2$ the interaction and kick Hamiltonians are
\begin{equation}
H_{I}= \sum_{n=1 }^N \left[ \frac{\pi}{4}\hat{\sigma}_{n}^{z} \hat{\sigma}_{n+2}^{z} + J_{2}(n)\hat{\sigma}_{n}^{z} \hat{\sigma}_{n+1}^{z} + d \hat{\sigma}_{n}^{z} \right], \qquad
H_{K}= \frac{\pi}{4}\sum_{ n=1}^N \hat{\sigma}_{n}^{x},
\end{equation} 
where, in the dual-unitary regime, parameters $J_1=h_x=\pi/4$ are fixed, while  the nearest-neighbor coupling $J_2(n)$ and the longitudinal field $d$ are arbitrary and may, in general, vary from site to site. We consider two particular cases. The first is a non-homogeneous kicked Ising chain (KIC), for which
\begin{equation}
J_2(n)=\left\{
\begin{array}{cc}
0, & n\; \mathrm{even},\\
G, & n\; \mathrm{odd},
\end{array}\right.
\end{equation}
while the second is the homogeneous KIC, in which $J_2(n)=G$ is uniform throughout the chain, and $d$ is fixed in both models. 

The local operators $\bm Q_1$ and $\bm Q_2$ entering the operators $\bar{\Sigma}$, and $\ubar{\Sigma}$ (eqs.~\ref{sigmaQ1},~\ref{sigmaQ2}) in our calculations of the correlation functions are chosen to have four point support: 
\begin{eqnarray}\label{bSigma}
\bm Q_1&=&\q_1\otimes \q_2\otimes \q_3\otimes \q_4;\\
\bm Q_2&=&\q_5\otimes \q_6\otimes\q_7\otimes \q_8,\label{Sigmab}
\end{eqnarray}
where each $\q_i$ acts on one-site two-dimensional Hilbert space, and they are chosen to satisfy $\tr \bm Q_1=\tr \bm Q_2 =0$.

\subsection{KIC with non-homogeneous interactions}\label{SecNonHom}
According to the discussion in the previous section, the non-homogeneous KIC corresponds to the following particular choice of the functions $g(s,\bar{s})$, $h(s,\bar{s})$, and functions $f_1(s,\bar{s})$ and $f_2(s,\bar{s})$ entering the functional $\mathcal{F}(s,\bar{s})$ in eq.~(\ref{Faction}): 
\begin{equation}\label{ghdualunitarychoice1}
    g(s,\bar{s})=
    \frac{\pi}{4}
    \left[
        (2s-3)(2\bar{s}-3)-1
    \right],\quad h(s,\bar{s})
    =g(s,\bar{s})+d(s-\bar{s}-3),
\end{equation}
\begin{equation}
f_1(s,\bar{s})=0,\qquad
f_2(s,\bar{s})=G(2s-3)(2\bar{s}-3).
\end{equation}
The choice of the coefficients $J$ and $h_x$ in $g(s,\bar{s})$ and $h(s,\bar{s})$ corresponds to the dual-unitary case, eq.~(\ref{ghdualunitarychoice}). The arguments $s$, and $\bar{s}$ take the values $1$ and $2$.  

The non-homogeneous KIC model implies that the number of particles $N$ is even. The correlation function calculation depends not only the particular choice of $\bm Q_1$, and $\bm Q_2$ but also on their position. We consider the case, when the operator $\q_1$ is placed on even site (see Fig.~\ref{Fig2App}), 
then the transfer operator $\hat{\bm T}$ obtained from eq.~(\ref{That16x16})  possesses a symmetric block-hierarchical structure. It can be represented as the tensor product
\begin{equation}\label{T4Ising}
    \hat{\bm T}=B\otimes E,
\end{equation}
where
\begin{equation}
B=
\left(
\begin{array}{cccc}
a & b & c & c\\
b & a & c & c\\
c & c & a & b\\
c & c & b & a
\end{array}
\right),
\qquad
E=\frac{1}{16}
\left(
\begin{array}{cccc}
1 & 1 & 1 & 1\\
1 & 1 & 1 & 1\\
1 & 1 & 1 & 1\\
1 & 1 & 1 & 1
\end{array}
\right),
\end{equation}
with $a$, $b$, and $c$ depending on the parameters $G$ and $d$,
\begin{eqnarray}
a &=& 1+\cos^2(2d)+2\cos(2d)\cos(2G),\\
b &=& 1+\cos^2(2d)-2\cos(2d)\cos(2G),\\
c &=& \sin^2(2d).
\end{eqnarray}
The transfer matrix eq.~(\ref{T4Ising}) has four non-zero eigenvalues. The largest eigenvalue is $\lambda_0=1$, with the eigenvector of identical elements
\begin{equation}\label{Eps0}
\E_0=
\frac{1}{2}
\sum_{\chi,\eta,\chi_1,\eta_1}
|\chi,\eta,\chi_1,\eta_1\rangle
=
\frac{1}{2}
e_0\otimes e_0\otimes e_0\otimes e_0.
\end{equation}
The eigenvalues 
$
\lambda_1=\lambda_2=\cos(2G)\cos(2d)
$
is doubly degenerate, with the corresponding eigenvectors
\begin{equation}\label{Eps1}
\E_1=
\frac{1}{2}
\sum_{\chi,\eta,\chi_1,\eta_1}
(-1)^{\chi+\eta}
|\chi,\eta,\chi_1,\eta_1\rangle
=
\frac{1}{2}
e_0\otimes e_1\otimes e_0\otimes e_0,
\end{equation}
and
\begin{equation}\label{Eps2}
\E_2=
\frac{1}{2}
\sum_{\chi,\eta,\chi_1,\eta_1}
(-1)^{\chi_1+\eta_1}
|\chi,\eta,\chi_1,\eta_1\rangle
=
\frac{1}{2}
e_1\otimes e_1\otimes e_0\otimes e_0.
\end{equation}
Finally,
$
\lambda_3=\cos^2(2d)$,
with the eigenvector
\begin{equation}\label{Eps3}
\E_3=
\frac{1}{2}
\sum_{\chi,\eta,\chi_1,\eta_1}
(-1)^{\chi+\eta+\chi_1+\eta_1}
|\chi,\eta,\chi_1,\eta_1\rangle
=
\frac{1}{2}
e_1\otimes e_0\otimes e_0\otimes e_0.
\end{equation}
Here,
$
e_0=|1\rangle+|2\rangle$,
 $
e_1=|1\rangle-|2\rangle$.
The vectors $\E_i$ form an orthonormal set,
$
\E_i^\dagger\E_j=\delta_{ij}.
$

The correlation function can be evaluated explicitly. Its form depends on the particular choice of the local operators $\q_i$ entering $\bar{\Sigma}$ and $\ubar{\Sigma}$ [see eqs.~(\ref{bSigma}) and~(\ref{Sigmab})]. Moreover, the choice of these operators determines the direction of correlation propagation, i.e., whether the relevant contribution propagates along the left or right edge of the light cone. Notice here that it is enough to calculate the correlation function for the operators $\q_i$ taken from the set of Pauli matrices $\set{\sigma^0,\sigma^x,\sigma^y,\sigma^z}$, which we supplemented by the two-dimensional identity matrix, denoted as $\sigma^0$.

For propagation along the left edge of the light cone, the correlation function takes the form
\begin{multline}\label{CexactNonuniformLeft}
C(t)=
-\Theta_I(\q_3,\q_4,\q_5,\q_6)A_I\big(\q_1,\q_2\big)
 A_I\big(\q_7,\q_8\big)
 \lambda_1^{t-2}
\\+
\Theta_{II}(\q_3,\q_4,\q_5,\q_6)A_{II}\big(\q_1,\q_8\big)
 A_{II}\big(\q_2,\q_7\big)
 \lambda_3^{t-2}.
\end{multline}
The factors $\Theta_I(\q_3,\q_4,\q_5,\q_6)$, and $\Theta_{II}(\q_3,\q_4,\q_5,\q_6)$ take values either $1$ or $0$, the function $\Theta_{II}(\q_3,\q_4,\q_5,\q_6)$ equals $1$ for traceless operators only, while $\Theta_{II}(\q_3,\q_4,\q_5,\q_6)$ equals $1$ when at least two of operators have trace $1$. The details of the calculation, together with all non-zero values of the amplitudes $A_I(\q_i,\q_j)$ and $A_{II}(\q_i,\q_j)$, are given in Appendix~\ref{AppBoundNonHom}. Below we reproduce the sets of $\q_i$ corresponding to non-trivial correlation functions.

The amplitude
$\Theta_I(\q_3,\q_4,\q_5,\q_6)A_I(\q_1,\q_2)A_I(\q_7,\q_8)$
is non-zero for the following combinations of local operators:
\begin{itemize}
    \item[(i)] $\q_3=\q_5=\sigma^0$, $\q_4=\sigma^z$, $\q_6=\sigma^y$, and
    $$
    \q_1\in\set{\sigma^0,\sigma^z},\qquad
    \q_2\in\set{\sigma^x,\sigma^y},\qquad
    \q_7\in\set{\sigma^0,\sigma^y},\qquad
    \q_8\in\set{\sigma^x,\sigma^z};
    $$

    \item[(ii)] $\q_4=\q_6=\sigma^0$, $\q_3=\sigma^z$, $\q_5=\sigma^y$, and
    $$
    \q_1\in\set{\sigma^x,\sigma^y},\qquad
    \q_2\in\set{\sigma^0,\sigma^z},\qquad
    \q_7\in\set{\sigma^x,\sigma^z},\qquad
    \q_8\in\set{\sigma^0,\sigma^y}.
    $$
\end{itemize}
The first combination of operators in (i) is associated with the eigenvector $\E_1$ [eq.~(\ref{Eps1})], whereas the second combination in (ii) corresponds to the eigenvector $\E_2$ [eq.~(\ref{Eps2})]. 

In contrast, the second term in eq.~(\ref{CexactNonuniformLeft}) vanishes for both sets (i) and (ii). It becomes non-zero when all local operators $\q_i$ are traceless. In particular, correlations propagating along the left edge of the light cone with the characteristic rate $\lambda_3$ occur for
\begin{itemize}
    \item[(iii)] $\q_3=\q_4=\sigma^z$, $\q_5=\q_6=\sigma^y$, and
    $$
    \q_1\in\set{\sigma^x,\sigma^y},\qquad
    \q_2\in\set{\sigma^x,\sigma^y},\qquad
    \q_7\in\set{\sigma^x,\sigma^z},\qquad
    \q_8\in\set{\sigma^x,\sigma^z}.
    $$
\end{itemize}

The corresponding combinations of operators for propagation along the right edge of the light cone can be obtained from spatial-reflection symmetry. The correlation function is invariant under spatial reversal, which induces the following re-enumeration of the local operators:
\begin{equation}\label{spatialinverse}
\q_1\leftrightarrow\q_3,\qquad
\q_2\leftrightarrow\q_4,\qquad
\q_5\leftrightarrow\q_7,\qquad
\q_6\leftrightarrow\q_8.
\end{equation}
Applying this transformation to the sets (i)--(iii) generates three additional sets of operators for which the correlation function is non-zero.

Finally, we note that eq.~(\ref{CexactNonuniformLeft}) reproduces the results known for Ising chains with nearest-neighbour interactions~\cite{GBAWG20}. In particular setting $G=0$ decouples ($N$ is even) the lattice into two non-interacting sublattices. Correlations then propagate independently along each sublattice. For example, for
$
\q_2=\q_4=\q_6=\q_8=\sigma^0,
$
corresponding to case (ii), only one of the two sublattices is ``activated.'' In this case, the eigenvalue $\lambda_2$ reduces to
$
\lambda_2=\cos(2d),
$
which coincides with the corresponding propagation rate for the Ising chain with nearest-neighbour interactions. Moreover, only the four configurations
$\q_3=\sigma^z$, $\q_5=\sigma^y$, and $\q_1=\set{\sigma^x,\sigma^y}$, $\q_7=\set{\sigma^x,\sigma^z}$
yield a non-trivial correlation function, in agreement with the known results~\cite{GBAWG20}.

The second term in eq.~(\ref{CexactNonuniformLeft}) is independent of the interaction strength $G$. It can therefore be interpreted as describing the independent propagation of correlations along the two decoupled sublattices. Since the propagation rate along each individual sublattice is $\cos(2d)$, the corresponding eigenvalue for the combined propagation is its square,
$\lambda_3=\cos^2(2d)$.

\subsection{KIC with homogeneous interactions}
\label{SecHom}

The homogeneous KIC with next-to-next-nearest-neighbor interactions differs from the inhomogeneous model considered above only in the choice of the functions $f_1$ and $f_2$. In the homogeneous case, we set
\begin{equation}
    f_1(s,\bar{s})
    =
    f_2(s,\bar{s})
    =
    G(2s-3)(2\bar{s}-3),
\end{equation}
while the functions $g(s,\bar{s})$ and $h(s,\bar{s})$ are chosen according to the dual-unitary condition, as specified in eq.~(\ref{ghdualunitarychoice1}). The corresponding functional $\mathcal{F}(s,\bar{s})$ is given by eq.~(\ref{Faction}).

The resulting transfer matrix is not symmetric, but its eigenvalues can nevertheless be obtained explicitly. Its four non-zero eigenvalues are
\begin{equation}
    \lambda_0=1,\qquad
    \lambda_1=\lambda_2=\cos(2d)\cos^2(2G),\qquad
    \lambda_3=\cos^2(2d)\cos^2(2G).
\end{equation}
The eigenvalue $\lambda_0=1$ corresponds to the constant eigenvector $\E_0$, given in eq.~(\ref{Eps0}). The remaining eigenvectors occur in pairs. Their explicit form depends on the particular ordering of the indices $\chi$, $\eta$, $\chi_1$, and $\eta_1$ used to construct the transfer matrix. The particular ordering of these indices, which we used for for calculations, together with the corresponding eigenvectors and further details of the calculation, is presented in Appendix~\ref{AppBoundHom}.

The correlation functions propagating along the left edge of the light cone have a structure similar to that obtained for the inhomogeneous case, see eq.~(\ref{CexactNonuniformLeft}). However, in the homogeneous model a larger set of operator combinations gives non-vanishing contributions. The corresponding correlations are governed by the eigenvalue $\lambda_1$ and are given explicitly by eqs.~(\ref{CexactUniformLeft1}) and~(\ref{CexactUniformLeft2}). They are non-vanishing for the following two classes of operator combinations:
\begin{itemize}
    \item[(i)] For
    $\q_4=\sigma^z$, $
        \q_5=\sigma^0$,
        $
        \q_6=\sigma^y$,
    with
    $$
        \q_1\in\{\sigma^0,\sigma^z\},\quad
        \q_2\in\{\sigma^x,\sigma^y\},\quad
        \q_3\in\{\sigma^0,\sigma^z\},
    \mbox{  and }\;\;
        \q_7\in\{\sigma^0,\sigma^y\},\quad
        \q_8\in\{\sigma^x,\sigma^z\}.
    $$

    \item[(ii)] For
    $        \q_3=\sigma^z$, $\q_4=\sigma^0$, $\q_5=\sigma^y$,
    with
    $$
        \q_1\in\{\sigma^x,\sigma^y\},\quad
        \q_2\in\{\sigma^0,\sigma^z\},\quad
        \q_6\in\{\sigma^0,\sigma^y\},
    \mbox{ and }\;\;
        \q_7\in\{\sigma^x,\sigma^z\},\quad
        \q_8\in\{\sigma^0,\sigma^y\}.
    $$
\end{itemize}

There is an additional contribution associated with the eigenvalue $\lambda_3$, which gives rise to a decay proportional to $\lambda_3^{t-2}$, see eq.~(\ref{CexactUniformRight3}). This contribution is non-vanishing for the following class of operator combinations:
\begin{itemize}
    \item[(iii)] For
    $\q_4=\sigma^z$, $\q_5=\sigma^y$,
    with
$$\q_1\in\{\sigma^x,\sigma^y\},\quad \q_2\in\{\sigma^x,\sigma^y\},\quad \q_3\in\{\sigma^0,\sigma^z\},
\mbox{    and }$$$$
         \q_6\in\{\sigma^0,\sigma^y\},\quad
        \q_7\in\{\sigma^x,\sigma^z\},\quad
        \q_8\in\{\sigma^x,\sigma^z\}.
 $$
\end{itemize}

The corresponding combinations of operators for propagation along the right edge of the light cone can be obtained from spatial-reflection symmetry, i.e. by re-enumeration of the local operators according to eq.~(\ref{spatialinverse}). Applying this transformation to the sets (i)--(iii) generates three additional sets of operators for which the correlation function is non-zero.

It is worth noting that, in contrast to the inhomogeneous case, the contribution proportional to $\lambda_3^{t-2}$ can survive even when one or both of the operators $\q_3$ and $\q_6$ are equal to $\sigma^0$, i.e., when these operators have non-zero trace.

Finally, in the limit $G=0$ for even number of particles $N$, the correlation functions reduce to those of the inhomogeneous model evaluated at $G=0$. The resulting expression is consistent with the previously known results~\cite{GBAWG20}.

\begin{figure}
\begin{center}
\begin{tabular}{lccc}
a)& \includegraphics[width=0.42\linewidth]{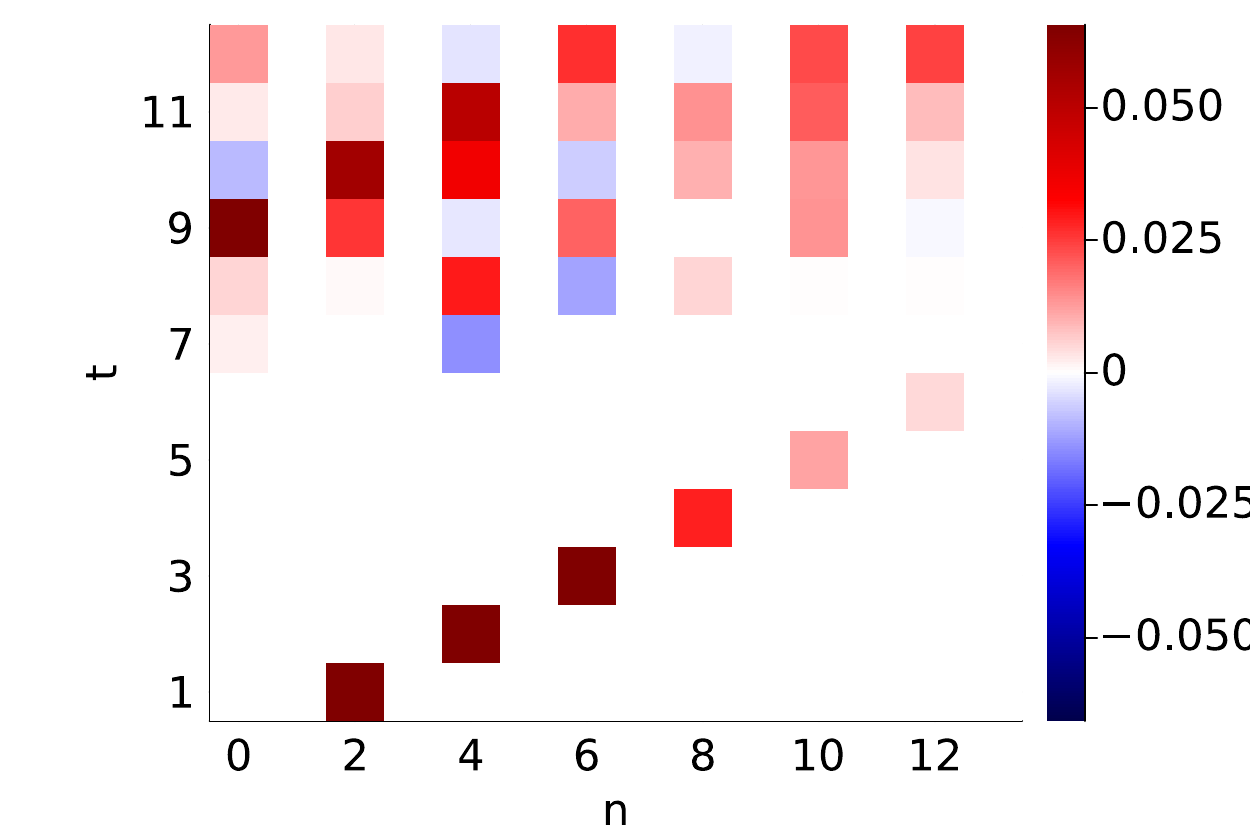}&b)&\includegraphics[width=0.42\linewidth]{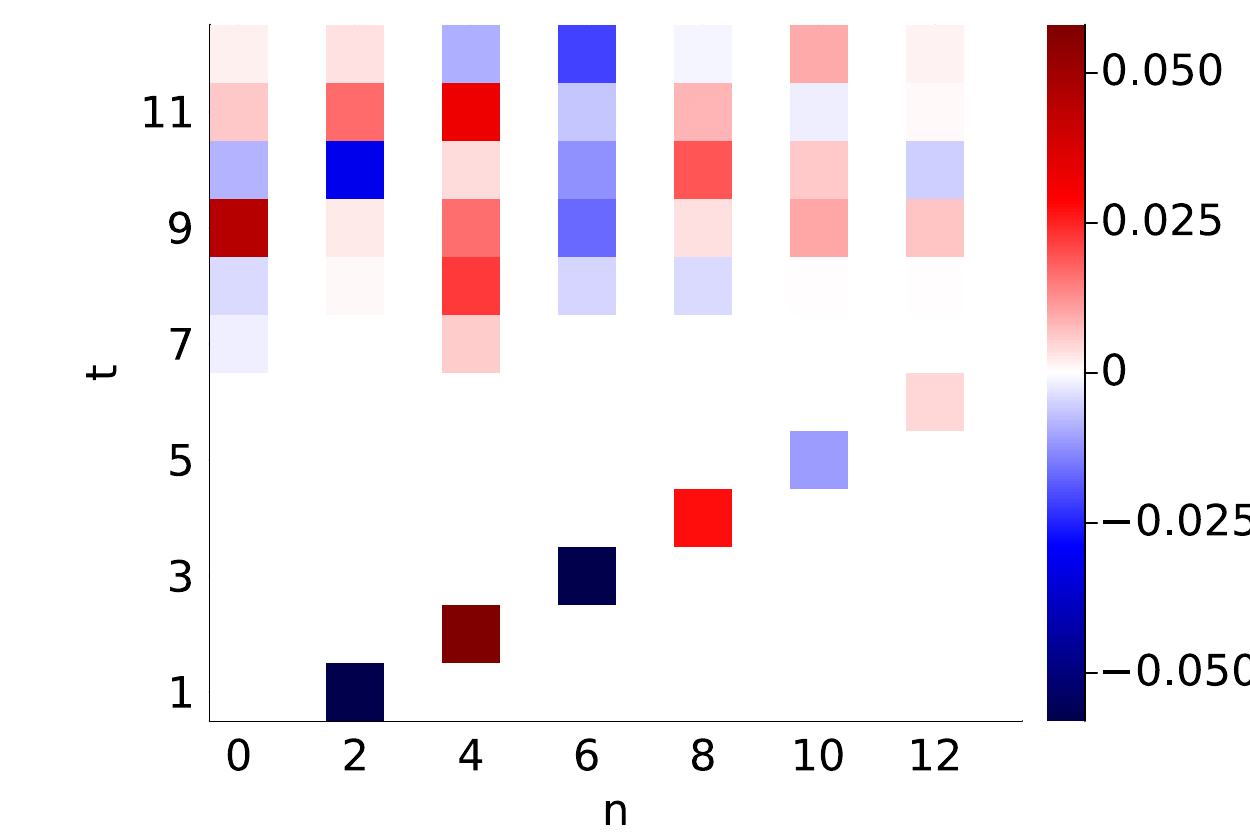}
\end{tabular}
\end{center}
\caption{\small
Two-point correlation functions between the operators defined in eqs.~(\ref{bSigma}) and (\ref{Sigmab}), for
$\q_1=\sigma^y$, $\q_2=\sigma^0$, $\q_3=\sigma^z$, $\q_4=\sigma^0$,
$\q_5=\sigma^y$, $\q_6=\sigma^0$, $\q_7=\sigma^z$, and $\q_8=\sigma^0$,
with $d=1$ and $G=1.5$, for the (a) non-homogeneous and (b) homogeneous models.
In both cases, the only nonvanishing correlations are those propagating along the light-cone edges, $|n|=2|t|$.
The ratios of consecutive nonzero correlation values agree with the corresponding subleading eigenvalue $\lambda_1$ of the transfer matrix. Specifically,
$
\frac{C(n=2(t+1),t+1)}{C(n=2t,t)}
=
\cos(2G)\cos(2d)
$
for the non-homogeneous model, whereas
$
\frac{C(n=2(t+1),t+1)}{C(n=2t,t)}
=
\cos^2(2G)\cos(2d)
$
for the homogeneous model.
}
\label{Bild13}
\end{figure}
\subsection{Numerical comparison}

To corroborate the analytical results, we have performed numerical calculations of the correlation function; the results are shown in Fig.~\ref{Bild13}. At the qualitative level, the numerical data are in agreement with the analytical predictions obtained above. In particular, nonvanishing correlations are observed only along the edges of the light cone, corresponding to
$
|n|=2|t|.
$
For a quantitative comparison, we consider the ratio of the nonzero correlation-function values at two consecutive time steps. The numerical results show that these ratios coincide with the corresponding subleading eigenvalues, $\lambda_{1,2}$, of the transfer matrices for the two models. Notice, that for large times, $t>N/2$ because of the cyclic boundary conditions, the correlation function looses its nice single-line structure. This means that in this regime the correlation are essentially non-local, and the proposed theory is not applicable to this case.

\subsection{Spectral statistics} 

\begin{figure} 
\begin{center}
\begin{tabular}{lccc}
a)& \includegraphics[width=0.5\linewidth]{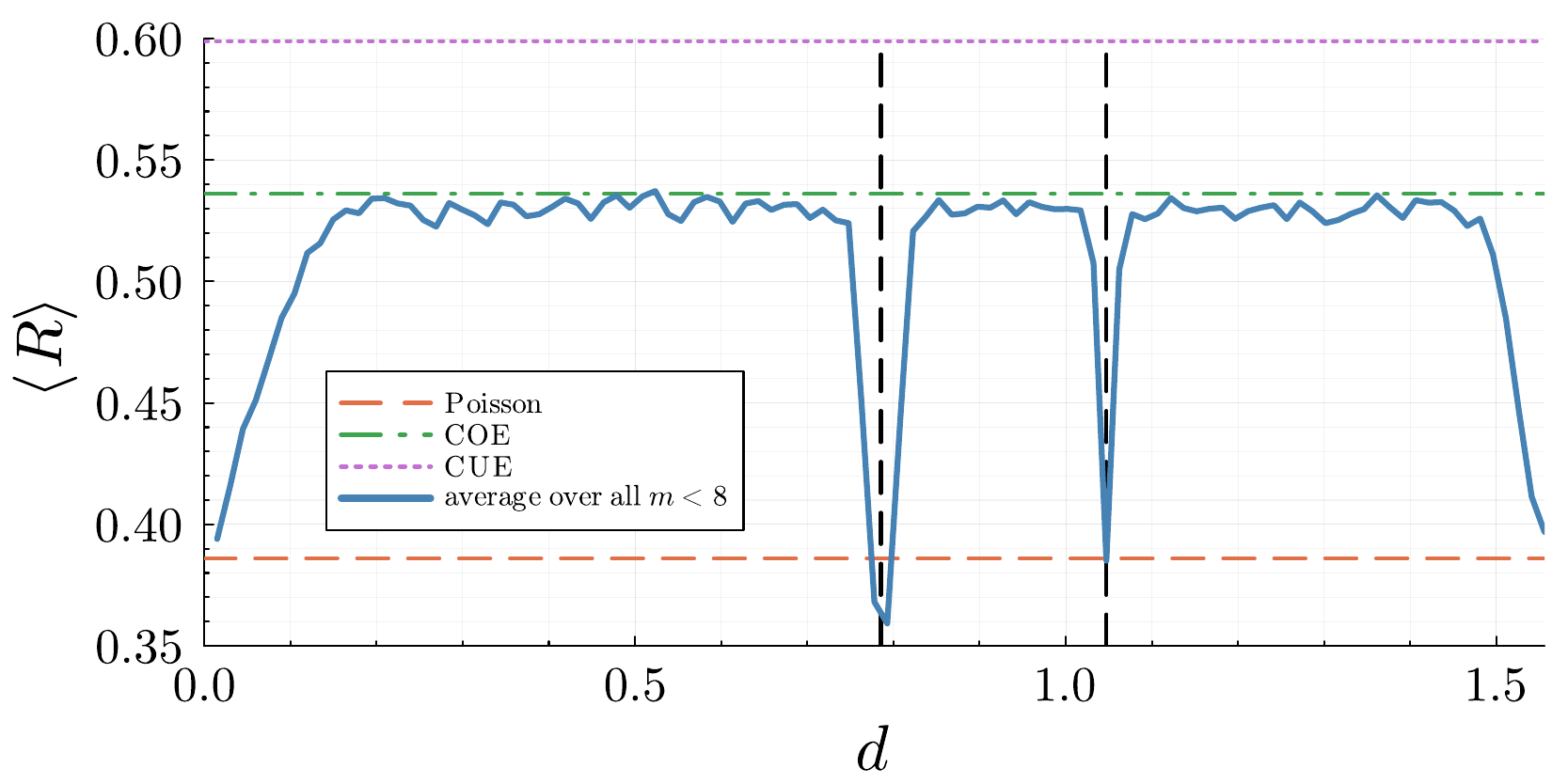}\\
b)&\includegraphics[width=0.5\linewidth]{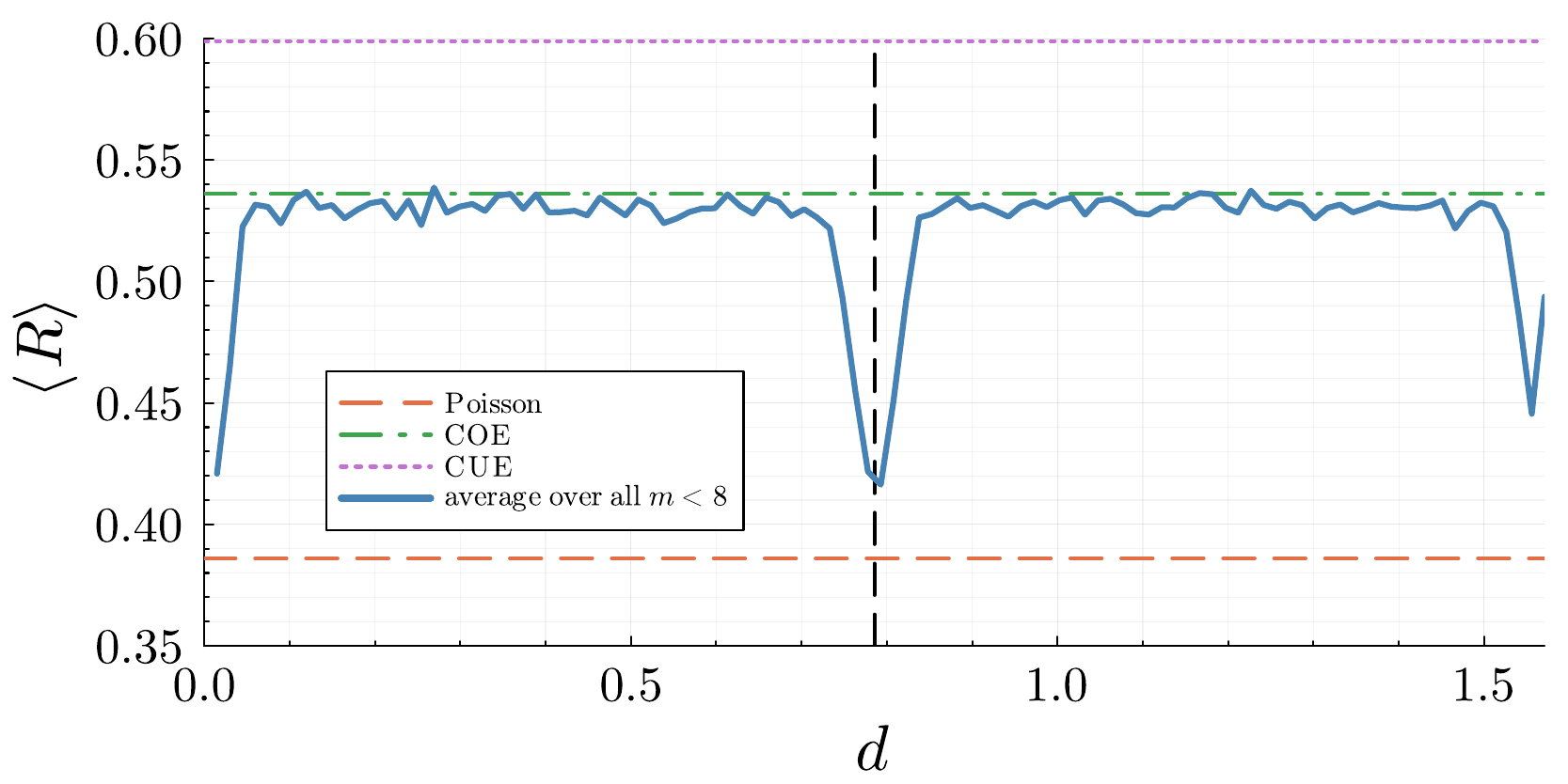}
\end{tabular}
\end{center}
\caption{\small The average value of  ratio of consecutive
level spacings eq.~(\ref{rvalue})
for the spectrum of the dual-unitary spin chain with the  next-to-next neighbor  interaction  for  $N=15$, $G=0$ (a) and $N=15$, $G=\pi/4$ (b), respectively. The value of $R$ is 
averaged over all values in the separate (translation) symmetry sectors of $k=1,\dots,7$ and shown as a function of $d$. Note that the graph is symmetric under reflection $d\to\pi- d$. For generic  values of $d$
the data for $\langle R\rangle$
fit the circular orthogonal ensemble (COE) prediction. The  dips at 
$d=0,\pi/4,\pi/3$ (a) and at  $d=0,\pi/4$ (b) correspond to spectral statistics characteristic of integrable systems.
}\label{Bild3}
\end{figure}
The exponential decay of the correlations in eq.~(\ref{CD1}) implies that the system exhibits quantum-chaotic behavior for generic values of the parameters \(G\) and \(d\). To investigate this issue in greater detail, it is instructive to examine the spectral statistics of the system.

Owing to translational symmetry in the model with homogeneous interactions, the spectrum of the spin-chain $2^{NT}$-dimensional evolution operator \(U\) (eq.~\ref{Ugeneric}) can be decomposed into \(N\) symmetry sectors, 
see~\cite{AWGG16}. The set of eigenvalues corresponding to the $n$-th sector ($n=1,2\dots,N$) we denote as $\set{e^{i\theta_k^{(n)}}}$, where $k=1,\dots,2^T$, and $T$ is the total number of time steps. In Fig.~\ref{Bild3}, we show the average ratio of consecutive level spacings within the same symmetry sector,
\begin{equation}
R=
\frac{
\min\left\{
\theta_k^{(n)}-\theta_{k-1}^{(n)},
\theta_{k+1}^{(n)}-\theta_k^{(n)}
\right\}
}{
\max\left\{
\theta_k^{(n)}-\theta_{k-1}^{(n)},
\theta_{k+1}^{(n)}-\theta_k^{(n)}
\right\}
}, \label{rvalue}  
\end{equation}
which is a well-established diagnostic of quantum chaos~\cite{OganesyanHuse2007, Bogomolny2013}.

For \(G=0\), the spectrum of the system coincides with that of the standard self-dual KIC with \(r=1\).   For generic values of \(d\), its spectral statistics agree with those of the circular orthogonal ensemble (COE) of random matrices. There are, however, several special values of \(d\) for which the KIC spectrum becomes ``non-chaotic''. The first two cases, \(d=0\) and \(d=\pi/4\), correspond to the well-known integrable points of the classical two-dimensional Ising model~\cite{LeeYangI,LeeYangII,Matveev_2008}. More intriguing is the case \(d=\pi/3\), where the spectrum displays  Poissonian level statistics despite the exponential decay of correlations, see Fig.~\ref{Bild3}a.

For the \(r=2\) kicked chain with homogeneous coupling \(G=\pi/4\), we obtain qualitatively similar results. Apart from the special points \(d=0\) and \(d=\pi/4\), the system exhibits COE spectral statistics, see Fig.~\ref{Bild3}b.

 \section{Conclusion}\label{Sec7}

 Models possessing space--time symmetry, such as the KIC, provide the most straightforward examples of systems with dual-unitary dynamics. In these systems, the dynamics is invariant under the exchange of space and time, which automatically implies unitary evolution in both the temporal and spatial directions. This symmetry has immediate consequences for the correlations between local operators. In particular, the two-point correlation functions \(C(n,t)\) must be symmetric under the interchange of space and time,
\(C(n,t)=C(t,n)\). Combined with causality, this symmetry implies that the connected correlation functions can be nonzero only along the light-cone edges \(|n|=|t|\).

For the kicked systems with exact space--time symmetry, only nearest-neighbor interactions are possible. Indeed, since the time evolution couples states at time \(t\) to states at time \(t-1\), 
the same local structure must also be preserved in the spatial direction.
In spite of this, we have shown in the present work that the interaction range can be extended to arbitrary \(r\) while preserving dual unitarity. Although the exact space--time symmetry is lost for \(r>1\), so that the temporal and spatial evolutions are generated by different operators, \(U\) and \(\tilde{U}\), respectively, both operators remain unitary. This, in turn, implies that correlations between local operators vanish only along the light cone edges  \(|n|=r|t|\), where they can be evaluated analytically. Beyond correlation functions, the present class of models also provides an effective framework for studying other aspects of many-body dynamics. For example, a recent preprint~\cite{TanayPathak2026} derived an exact formula for the entanglement dynamics in a subclass of dual-unitary spin chains with long-range interactions.

The construction we propose yields a very broad family of dual-unitary models. For a fixed interaction between spins separated by a distance $r$, the general interactions involving the remaining $r-1$ neighboring spins can be chosen arbitrarily. In particular, by introducing an additional, precisely tuned pairwise interaction between spins separated by $r-1$ sites, one can impose dual unitarity between them as well. Since the number of possible interaction terms grows rapidly with \(r\), this construction generates a broad and highly flexible class of dual-unitary spin chains. 
Such flexibility may prove valuable for investigating the spectral properties of these systems. For example, the proof of universal spectral statistics for a certain class of spin chains presented in Ref.~\cite{KosLjubotinaProsen2018} relies crucially on the presence of long-range interactions 
and on the ability to perform averages over them. It would therefore be interesting to investigate whether  a similar approach can be applied to the long-range dual-unitary spin chains introduced in the present work.

\section*{Acknowledgements}
During the preparation of this manuscript, we became aware of the independent work on entanglement dynamics in a related model. We are grateful to Tanay Pathak for sharing with us a preliminary version of his preprint~\cite{TanayPathak2026}. 



\begin{appendix}
\numberwithin{equation}{section}

\section{Proof of Proposition 1}\label{AppA}

The left-hand side of the duality relation (\ref{duality}) can be written as
\begin{eqnarray}&&\trace \,  {U}^T=\prod_{t=1}^T\langle\bm s_t|U| \bm s_{t+1}\rangle= \nonumber\\
&&=\left(\frac{1}{L}\right)^{\frac{NT}{2}}\sum_{\{s_{n,t}\}}\prod_{n=1}^N\prod_{t=1}^Te^{
i \left(g(s_{n,t},s_{n,t+1}) +h(s_{n,t},s_{n+r,t}) + f_n(s_{n,t},\dots,s_{n+r-1,t}) 
\right)},\label{QQduality}
\end{eqnarray}
where we assume periodic boundary conditions, $s_{n,t}\equiv s_{n\, \mathrm{ mod }N,t\,\mathrm{ mod } T}$.

To prove Proposition 1, it remains to show that the right-hand side of eq.~(\ref{duality}) yields the same expression. Expanding the trace in the product basis, we obtain 
\begin{eqnarray}&&\trace \,  \prod_{m=1}^\N \tilde{U}_m=\prod_{m=1}^{\N}\langle\bm \eta_{m-1}|\tilde{U}_m| \bm \eta_{m}\rangle=\prod_{m=1}^\N\langle\bm \eta_{m-1}|
\otimes_{t=1}^{T } \tilde{u}[ h,\bm f_m] U_I[g]| \bm \eta_{m}\rangle\\\nonumber
&& = \prod_{m=0}^{\N-1}\prod_{t=1}^{T }\langle\bm \eta_{m,t}|
 \tilde{u}[ h,\bm f_m]| \bm \eta_{m+1,t}\rangle e^{
i \sum_{j=1}^r g(\eta_{mr+j,t},\eta_{mr+j,t+1})}
, \label{QQduality1}
\end{eqnarray}
where the basis states are defined as 
\begin{equation} \set{\ket{\bm \eta_m}\equiv\otimes_{t=1}^T \ket{\bm\eta_{m, t}} , \,\,\, \ket{\bm \eta_{m,t}}\equiv\otimes_{j=1}^r \ket{\eta^t_{mr+ j}} |, \,\,\,  \eta^t_{mr+j}=\overline{1,L}}. \label{basisdual2}\end{equation}
By using eq.~(\ref{dualop}) we obtain 
\begin{equation}\trace \,  \prod_{m=1}^\N \tilde{U}_m=\left(\frac{1}{L}\right)^{\frac{NT}{2}}\sum_{\{\eta^t_{m,t}\}}\prod_{m=0}^{\N-1}\prod_{j=1}^r\prod_{t=1}^T
e^{i  \mathcal{S}(\{\eta^t_{m}\})},\label{QQduality3}
\end{equation} 
where
\begin{eqnarray}
&&\mathcal{S}(\{\eta^t_{m}\})= g(\eta^t_{mr+j},\eta^{t+1}_{mr+j})+ h(\eta^t_{mr+j},\eta^t_{(m+1)r+j})+\nonumber\\
&&+f_{mr+j}(\eta^t_{mr+j},\cdots,\eta^t_{mr+r}, \eta^t_{mr+r+1},\cdots,\eta^t_{mr+r+j-1} ).  
\end{eqnarray}
Finally, substituting $n=mr+j$ and relabeling the variables $\eta^t_{mr+j}\to s_{n,t}$, we recover exactly the expression in eq.~(\ref{QQduality}). This proves the duality relation~(\ref{duality}).

\section{Calculation of the correlation function. Non-homogeneous KIC}
\label{AppBoundNonHom}

To calculate the correlation function in eq.~(\ref{CD1}), one needs,
in addition to the transfer matrix, its eigenvalues and corresponding
eigenvectors, to construct the boundary vectors
$\bra{\bar{\bm\Phi}_{\bar{\Sigma}}}$ and
$\ket{\bm \Phi_{\ubar{\Sigma}}}$.
For the operators $\bar{\Sigma}$ and $\ubar{\Sigma}$ with four-point
supports (see eqs.~\ref{bSigma} and~\ref{Sigmab}), these vectors are
determined by the sets of local operators
$\q_1,\q_2,\q_3,\q_4$ and
$\q_5,\q_6,\q_7,\q_8$, respectively.

We consider the propagation of correlations along the left edge of
the light cone. In this case, the components of the vector
$\bra{\bar{\bm\Phi}_{\bar{\Sigma}}}$ are given by
\begin{eqnarray}\label{Phi1}
\braket{\bar{\bm\Phi}_{\bar{\Sigma}}|\chi,\eta,\chi_1,\eta_1}
&=&
\frac{1}{2^7}
\sum_{s_1,\bar{s}_3,\ubar{s}_3}
\sum_{s_2,\bar{s}_4,\ubar{s}_4}
\bar{\Gamma}_{s_3,s_4}^{s_1,\bar{s}_1,s_2,\bar{s}_2}
(\chi,\eta,\chi_1,\eta_1)
\nonumber\\
&&\times
\bra{s_1}\q_1\ket{\bar{s}_1}
\bra{s_2}\q_2\ket{\bar{s}_2}
\bra{s_3}\q_3\ket{s_3}
\bra{s_4}\q_4\ket{s_4}.
\end{eqnarray}

The coefficient
$\bar{\Gamma}_{s_3,s_4}^{s_1,\bar{s}_1,s_2,\bar{s}_2}
(\chi,\eta,\chi_1,\eta_1)$
can be read off directly from the tensor-network representation
shown in Fig.~\ref{Fig2App}. It is given by
\begin{eqnarray}\label{PhiG1}
\bar{\Gamma}_{s_3,s_4}^{s_1,\bar{s}_1,s_2,\bar{s}_2}
(\chi,\eta,\chi_1,\eta_1)
&=&
e^{- i g(\chi,s_1) + i g(\chi,\bar{s}_1)
   - i g(\chi_1,s_2) + i g(\chi_1,\bar{s}_2)}
\nonumber\\
&&\times
e^{- i h(s_1,s_3) + i h(\bar{s}_1,s_3)
   - i h(\eta,s_1) + i h(\bar{s}_1,\eta)}
\nonumber\\
&&\times
e^{- i h(s_2,s_4) + i h(\bar{s}_2,s_4)
   - i h(\eta_1,s_2) + i h(\bar{s}_2,\eta_1)}
\nonumber\\
&&\times
e^{- i f_1(s_2,s_3) + i f_1(\bar{s}_2,s_3)
   - i f_1(s_1,\eta_1) + i f_1(\bar{s}_1,\eta_1)}
\nonumber\\
&&\times
e^{- i f_2(s_1,s_2)+i f_2(\bar{s}_1,\bar{s}_2)}.
\end{eqnarray}

The vector $\ket{\bm\Phi_{\ubar{\Sigma}}}$ can be constructed in an
analogous manner. It contains, however, one additional time slice
(see, e.g., the calculations in
Refs.~\cite{GBAWG20,GutOsipovGuhr2024}).
Consequently, its components are given by
\begin{eqnarray}\label{Phi2}
\braket{\chi',\eta',\chi'_1,\eta'_1|\bm\Phi_{\ubar{\Sigma}}}
&=&
\frac{1}{2^7}
\sum_{\bar{s}_1,\ubar{s}_1,s_3}
\sum_{\bar{s}_2,\ubar{s}_2,s_4}
\Gamma_{s_1,s_2}^{s_3,\bar{s}_3,s_4,\bar{s}_4}
(\eta',\chi',\eta'_1,\chi'_1)
\nonumber\\
&&\times
\bra{s_1}u^\dag[g]\q_5u[g]\ket{s_1}
\bra{s_2}u^\dag[g]\q_6u[g]\ket{s_2}
\nonumber\\
&&\times
\bra{s_3}u^\dag[g]\q_7u[g]\ket{\bar{s}_3}
\bra{s_4}u^\dag[g]\q_8u[g]\ket{\bar{s}_4}.
\end{eqnarray}

Here, the coefficient
$\Gamma_{s_1,s_2}^{s_3,\bar{s}_3,s_4,\bar{s}_4}
(\eta',\chi',\eta'_1,\chi'_1)$
is given by
\begin{eqnarray}\label{PhiG2}
\Gamma_{s_1,s_2}^{s_3,\bar{s}_3,s_4,\bar{s}_4}
(\eta',\chi',\eta'_1,\chi'_1)
&=&
e^{- i g(s_3,\eta') + i g(\bar{s}_3,\eta')
   - i g(s_4,\eta'_1) + i g(\bar{s}_4,\eta'_1)}
\nonumber\\
&&\times
e^{- i h(s_1,s_3) + i h(s_1,\bar{s}_3)
   - i h(s_3,\chi') + i h(\bar{s}_3,\chi')}
\nonumber\\
&&\times
e^{- i h(s_2,s_4) + i h(s_2,\bar{s}_4)
   - i h(s_4,\chi'_1) + i h(\chi'_1,\bar{s}_4)}
\nonumber\\
&&\times
e^{- i f_1(s_2,s_3) + i f_1(s_2,\bar{s}_3)
   - i f_1(s_4,\chi') + i f_1(\bar{s}_4,\chi')}
\nonumber\\
&&\times
e^{- i f_2(s_3,s_4)+i f_2(\bar{s}_3,\bar{s}_4)}.
\end{eqnarray}

\begin{figure}
\begin{tikzpicture}
     \def\xstep{2.3}
     \def\arrowlen{0.4}
     \def\ypos{0}
     \def\arcHeight{0.8}
     \def\yshift{0.7}

     \foreach \i/\j in {1/3, 3/5} {
         \pgfmathsetmacro{\xa}{\i*\xstep}
         \pgfmathsetmacro{\xb}{\j*\xstep}
         \draw[red, very thick] (\xa,\ypos+\yshift) to (\xb,\ypos+\yshift);
     }

     \foreach \i/\j in {0/1, 2/3, 4/5} {
         \pgfmathsetmacro{\xa}{\i*\xstep}
         \pgfmathsetmacro{\xb}{\j*\xstep}
         \draw[blue, very thick] (\xa,\ypos) -- (\xb,\ypos+\yshift);
     }

     \foreach \i/\j in {1/2, 3/4} {
         \pgfmathsetmacro{\xa}{\i*\xstep}
         \pgfmathsetmacro{\xb}{\j*\xstep}
         \draw[yellow, very thick] (\xa,\ypos+\yshift) -- (\xb,\ypos);
     }

     \draw[green, very thick] (2*\xstep,\ypos) --
         (2*\xstep,\ypos-\xstep);

     \draw[green, very thick] (3*\xstep,\ypos+\yshift) --
         (3*\xstep,\ypos-\xstep+\yshift);

     \foreach \i/\j in {0/2, 2/4} {
         \pgfmathsetmacro{\xa}{\i*\xstep}
         \pgfmathsetmacro{\xb}{\j*\xstep}
         \draw[red, very thick] (\xa,\ypos) to (\xb,\ypos);
     }

     \filldraw[color=black, fill=white, very thick]
         (2*\xstep,\ypos) ellipse (30pt and 10pt);
     \node at (2*\xstep,\ypos)
         {\small $\bra{s_1}\q_1\ket{\bar{s}_1}$};

     \filldraw[color=black, fill=white, very thick]
         (3*\xstep,\ypos+\yshift) ellipse (30pt and 10pt);
     \node at (3*\xstep,\ypos+\yshift)
         {\small $\bra{s_2}\q_2\ket{\bar{s}_2}$};

     \filldraw[color=black, fill=white, very thick]
         (4*\xstep,\ypos) ellipse (30pt and 10pt);
     \node at (4*\xstep,\ypos)
         {\small $\bra{s_3}\q_3\ket{s_3}$};

     \filldraw[color=black, fill=white, very thick]
         (5*\xstep,\ypos+\yshift) ellipse (30pt and 10pt);
     \node at (5*\xstep,\ypos+\yshift)
         {\small $\bra{s_4}\q_4\ket{s_4}$};

     \filldraw[color=black, fill=white, very thick]
         (0*\xstep,\ypos) circle (8pt);
     \filldraw[color=black, fill=white, very thick]
         (1*\xstep,\ypos+\yshift) circle (8pt);

     \filldraw[color=black, fill=white, very thick]
         (2*\xstep,\ypos-\xstep) circle (8pt);
     \filldraw[color=black, fill=white, very thick]
         (3*\xstep,\ypos-\xstep+\yshift) circle (8pt);

     \node at (0*\xstep, \ypos) {\small $\eta$};
     \node at (1*\xstep, \ypos+\yshift) {\small $\eta_1$};

     \node at (2*\xstep, \ypos-\xstep) {\small $\chi$};
     \node at (3*\xstep, \ypos-\xstep+\yshift) {\small $\chi_1$};

\end{tikzpicture}
\caption{\small
Scheme illustrating the calculation of the boundary vector
$\braket{\bar{\bm\Phi}_{\bar{\Sigma}}|\chi,\eta,\chi_1,\eta_1}$
in eq.~(\ref{Phi1}). The colors of the edges are the same as in
Fig.~\ref{Fig1}.
}
\label{Fig2App}
\end{figure}

To obtain the correlation function, we diagonalize the transfer
operator and evaluate the projections of the boundary vectors
$\bra{\bar{\bm\Phi}_{\bar{\Sigma}}}$ and
$\ket{\bm\Phi_{\ubar{\Sigma}}}$ onto the eigenvectors
$\E_i$ defined in eqs.~(\ref{Eps0})--(\ref{Eps3}).

The projection onto $\E_0$ gives, as expected, a product of the
traces of all local operators $\q_i$. Thus, if at least one of the
operators $\q_i$ is traceless, this contribution vanishes and can be
discarded.

The contribution proportional to
$\lambda_{1,2}=\cos(2G)\cos(2d)$ in eq.~(\ref{CexactNonuniformLeft})
originates from the projections onto the eigenvectors $\E_1$ and
$\E_2$ (see eqs.~\ref{Eps1} and \ref{Eps2}). For propagation along
the left edge of the light cone, these projections are proportional
to the traces of selected local operators:
\begin{equation}
\bra{\bar{\bm\Phi}_{\bar{\Sigma}}}\cdot\E_1
\propto \tr \q_3,\qquad
\E_1\cdot\ket{\bm\Phi_{\ubar{\Sigma}}}
\propto \tr \q_5,
\qquad
\bra{\bar{\bm\Phi}_{\bar{\Sigma}}}\cdot\E_2
\propto \tr \q_4,\qquad
\E_2\cdot\ket{\bm\Phi_{\ubar{\Sigma}}}
\propto \tr \q_6.
\end{equation}
We obtain the following expression for the correlation function:
\begin{equation}\label{CexactNonuniformLeftApp}
C(t)=
-A_I\Big(\q_1,\q_2\Big)
 A_I\Big(\q_7,\q_8\Big)
\lambda_1^{t-2}.
\end{equation}
Here, the prefactor $A_I(\q_i,\q_j)$ is given in
Table~\ref{Table1}.

The correlation function is nonzero for propagation along the left
edge of the light cone in the following cases:
\begin{itemize}
    \item $\q_3=\q_5=\sigma^0$, $\q_4=\sigma^z$,
    $\q_6=\sigma^y$, with
    $\q_1=\set{\sigma^0,\sigma^z}$,
    $\q_2=\set{\sigma^x,\sigma^y}$,
    $\q_7=\set{\sigma^0,\sigma^y}$, and
    $\q_8=\set{\sigma^x,\sigma^z}$.
\end{itemize}
This set of operators corresponds to the eigenvector $\E_1$.
For the eigenvector $\E_2$, the correlation function is nonzero for
\begin{itemize}
    \item $\q_4=\q_6=\sigma^0$, $\q_3=\sigma^z$,
    $\q_5=\sigma^y$, with
    $\q_1=\set{\sigma^x,\sigma^y}$,
    $\q_2=\set{\sigma^0,\sigma^z}$,
    $\q_8=\set{\sigma^0,\sigma^y}$, and
    $\q_7=\set{\sigma^x,\sigma^z}$.
\end{itemize}

\begin{table}
\begin{center}
\begin{tabular}{ c|ccc}
\hline
$A_I\Big(\q_i,\q_j\Big)$
& $\q_i=\sigma^0$
& $\q_i=\sigma^y$
& $\q_i=\sigma^z$ \\
\hline
$\q_j=\sigma^x$
& $\sin(2d)\cos(2G)$
& $-\cos(2d)\sin(2G)$
& $\cos(2d)\sin(2G)$ \\
$\q_j=\sigma^y$
& $\cos(2d)\cos(2G)$
& 0
& $-\sin(2d)\sin(2G)$ \\
$\q_j=\sigma^z$
& $\cos(2d)\cos(2G)$
& $\sin(2d)\sin(2G)$
& 0
\\
\hline
\end{tabular}
\end{center}
\caption{\label{Table1}\small
Values of the prefactor $A_I(\q_i,\q_j)$ entering
eqs.~(\ref{CexactNonuniformLeftApp}, \ref{BIdef}).
}
\end{table}

The contribution proportional to
$\lambda_3^{t-2}$ in eq.~(\ref{CexactNonuniformLeft}), where
$\lambda_3=\cos^2 2d$, originates from the projection onto the
eigenvector $\E_3$ (see eq.~\ref{Eps3}). It is given by
\begin{equation}\label{CexactNonuniform3L}
C(t)=
A_{II}\Big(\q_1,\q_8\Big)
A_{II}\Big(\q_2,\q_7\Big)
\lambda_3^{t-2}.
\end{equation}
The values of the prefactor
$A_{II}(\q_i,\q_j)$ are given in Table~\ref{Table11}.
This contribution is nonzero when all local operators $\q_i$ are
traceless, namely for the following combinations:
\begin{itemize}
    \item $\q_3=\q_4=\sigma^z$,
    $\q_5=\q_6=\sigma^y$, with
    $\q_1=\set{\sigma^x,\sigma^y}$,
    $\q_2=\set{\sigma^x,\sigma^y}$,
    $\q_7=\set{\sigma^x,\sigma^z}$, and
    $\q_8=\set{\sigma^x,\sigma^z}$.
\end{itemize}

The operator combinations contributing to the correlation function
propagating along the left edge, as well as the corresponding forms
of the correlation function, can also be obtained from symmetry
considerations. In particular, spatial reflection symmetry induces
the following re-enumeration of the local operators:
\begin{equation}\label{re-enumeration}
\q_1\leftrightarrow\q_3,\qquad
\q_2\leftrightarrow\q_4,\qquad
\q_5\leftrightarrow\q_7,\qquad
\q_6\leftrightarrow\q_8.
\end{equation}

\begin{table}
\begin{center}
\begin{tabular}{ c|ccc}
\hline
$A_{II}\Big(\q_i,\q_j\Big)$
& $\q_i=\sigma^x$
& $\q_i=\sigma^y$
& $\q_i=\sigma^z$ \\
\hline
$\q_j=\sigma^x$
& $\sin^2(2d)$
& $\sin(2d)\cos(2d)$
& $\sin(2d)\cos(2d)$ \\
$\q_j=\sigma^y$
& $\sin(2d)\cos(2d)$
& $\cos^2(2d)$
& 0\\
$\q_j=\sigma^z$
& $\sin(2d)\cos(2d)$
& $\cos^2(2d)$
& $\cos^2(2d)$ \\
\hline
\end{tabular}
\end{center}
\caption{\label{Table11}\small
Values of the prefactor $A_{II}(\q_i,\q_j)$ entering
eqs.~(\ref{CexactNonuniform3L}, ~\ref{BIIIdef}).
}
\end{table}

\section{Calculation of the correlation function. Homogeneous KIC}\label{AppBoundHom}

For the homogeneous case, Sec.~\ref{SecHom}, the boundary
vectors and the explicit expressions for the correlation function can
be obtained in a manner analogous to the non-homogeneous case discussed
in Sec.~\ref{AppBoundNonHom}. The main difference is that the
homogeneous transfer matrix is asymmetric, requiring separate
calculations of its left and right eigenvectors. Here, we present the
results for correlation functions propagating along the left edge of
the light cone, using the expressions in eqs.~(\ref{Phi1}--\ref{PhiG2}).
The corresponding correlation functions propagating along the right
edge can be obtained using the re-enumeration rule formulated in
eq.~(\ref{re-enumeration}).

The explicit form of the transfer matrix depends on the ordering of
the indices $\chi$, $\eta$, $\chi_1$, and $\eta_1$. In our calculations,
we use the following ordering:
\begin{equation}
    \begin{array}{ccccccccccccccccc}
       \mathrm{Num.}: & 1\: & 2\: & 3\: & 4\: & 5\: & 6\: & 7\: & 8\: &
       9\: & 10 & 11 & 12 & 13 & 14 & 15 & 16 \\\hline
     \chi :  & 1 & 2 & 1 & 2 & 2 & 1 & 1 & 2 & 2 & 1 & 2 & 1 & 1 & 2 & 2 & 1 \\
     \eta  :  & 1 & 1 & 1 & 1 & 2 & 2 & 2 & 2 & 1 & 1 & 1 & 1 & 2 & 2 & 2 & 2 \\
     \chi_1: & 1 & 1 & 1 & 1 & 2 & 2 & 2 & 2 & 2 & 2 & 2 & 2 & 1 & 1 & 1 & 1 \\
     \eta_1: & 1 & 2 & 2 & 1 & 2 & 1 & 2 & 1 & 2 & 1 & 1 & 2 & 1 & 2 & 1 & 2
    \end{array}
\end{equation}
Each column of the table specifies the combination of indices
corresponding to the respective row or column number of the transfer
matrix, denoted by $\mathrm{Num.}$

The degenerate eigenvalue
$
\lambda_1=\lambda_2=\cos(2d)\cos^2(2G)
$
corresponds to two pairs of left and right eigenvectors:
\begin{equation}\label{Eps1inh}
    \tilde{\E}_{1}^{(L)}
    =\frac{1}{2}e_0\otimes e_1\otimes e_0 \otimes e_0,\quad
    \tilde{\E}_{1}^{(R)}
    =e_1\otimes
    \left(
    \begin{array}{c}
         -\sigma^0\psi\\
         -\sigma^x\psi\\
          \sigma^x\psi\\
          \sigma^0\psi
    \end{array}
    \right)
    ,
\end{equation}
\begin{equation}\label{Eps2inh}
    \tilde{\E}_{2}^{(L)}
    =e_1\otimes
    \left(
    \begin{array}{c}
         \sigma^0\psi\\
         \sigma^x\psi
    \end{array}
    \right)
    \otimes e_0,\quad
    \tilde{\E}_{2}^{(R)}
    =\frac{1}{2}e_1\otimes e_0\otimes e_0 \otimes e_0,
\end{equation}
with the vector $\psi$ given by
\begin{equation}
\psi= \frac{1}{8\cos(2d)\,\cos(2G)}\left(
    \begin{array}{c}
         \cos(2d-2G)
         \\[2mm]
         \cos(2d+2G)
    \end{array}
    \right).
\end{equation}
The eigenvalue
$
\lambda_3 = \cos^2(2d) \cos^2(2G)
$ corresponds to the left and right eigenvectors
$\tilde{\E}_3^{(L)}$ and $\tilde{\E}_3^{(R)}$ given by
\begin{equation}\label{Eps33}
 \tilde{\E}_3^{(L)}
 =e_0\otimes
 \left(
 \begin{array}{c}
     -\sigma^0\psi\\
      \sigma^x\psi
 \end{array}
 \right)
 \otimes e_0,
 \quad
 \tilde{\E}_3^{(R)}
 =e_0\otimes
 \left(
 \begin{array}{c}
     -\sigma^0\psi\\
     -\sigma^x\psi\\
      \sigma^x\psi\\
      \sigma^0\psi
 \end{array}
 \right).
\end{equation}

The left and right eigenvectors are normalized such that, together with
the eigenvector $\E_0$, they form a biorthonormal set,
${\tilde{\E}_i^{(L)}}{}^\dag \tilde{\E}_j^{(R)}
=\delta_{ij}$.

The eigenvectors above allow us to evaluate the correlation function.
For propagation along the left edge of the light cone, the correlation
function is nonzero for the following two configurations of operators $\q_i$:
\begin{itemize}
    \item For
    $\q_4=\sigma^z$, $\q_5=\sigma^0$, $\q_6=\sigma^y$, with
    $\q_1=\set{\sigma^0,\sigma^z}$,
    $\q_2=\set{\sigma^x,\sigma^y}$,
    $\q_3=\set{\sigma^0,\sigma^z}$,
    $\q_7=\set{\sigma^0,\sigma^y}$, and
    $\q_8=\set{\sigma^x,\sigma^z}$,
    the correlation function is
\begin{equation}\label{CexactUniformLeft1}
C(t)=
\cos(2G)
B_{I}\Big(\q_1,\q_2,\q_3\Big)
A_{I}\Big(\q_7,\q_8\Big)
\lambda_1^{t-2}.
\end{equation}

    \item For
    $\q_3=\sigma^z$, $\q_4=\sigma^0$, $\q_5=\sigma^y$, with
    $\q_1=\set{\sigma^x,\sigma^y}$,
    $\q_2=\set{\sigma^0,\sigma^z}$,
    $\q_6=\set{\sigma^0,\sigma^y}$,
    $\q_7=\set{\sigma^x,\sigma^z}$, and
    $\q_8=\set{\sigma^0,\sigma^y}$,
    the correlation function is
\begin{equation}\label{CexactUniformLeft2}
C(t)=
-\cos(2G)
A_{I}\Big(\q_2,\q_1\Big)
B_{I}\Big(\q_6,\q_7,\q_8\Big)
\lambda_1^{t-2}.
\end{equation}
\end{itemize}
Here,
\begin{equation}\label{BIdef}
B_{I}\Big(\q_i,\q_j,\q_k\Big)=
\left\{
\begin{array}{ccl}
\cos(2G) A_{I}\Big(\q_i,\q_j\Big),
&& \q_k=\sigma^0,\\[1mm]
-\sin(2G) B_{II}\Big(\q_i,\q_j\Big),
&& \q_k=\sigma^z,\;\sigma^y.
\end{array}
\right.
\end{equation}
The values of the prefactor $A_I(\q_i,\q_j)$ entering
eq.~(\ref{CexactUniformLeft2}) are given in Table~\ref{Table1},
while the values of $B_{II}(\q_i,\q_j)$ entering eq.~(\ref{BIdef})
are listed in Table~\ref{Table2}.

\begin{table}
\begin{center}
\begin{tabular}{ c|ccc}
 \hline
$B_{II}\Big(\q_i,\q_j\Big)$
& $\q_i=\sigma^0$
& $\q_i=\sigma^y$
& $\q_i=\sigma^z$ \\
\hline
 $\q_j=\sigma^x$
 & $-\sin(2d)\cos(2G)$
 & $-\cos(2d)\sin(2G)$
 & $\cos(2d)\sin(2G)$ \\
 $\q_j=\sigma^y$
 & $\cos(2d)\cos(2G)$
 & 0
 & $\sin(2d)\sin(2G)$ \\
 $\q_j=\sigma^z$
 & $\cos(2d)\cos(2G)$
 & $-\sin(2d)\sin(2G)$
 & 0
 \\
 \hline
\end{tabular}
\end{center}
\caption{\label{Table2}\small
Values of the prefactor $B_{II}(\q_i,\q_j)$ entering
eq.~(\ref{BIdef}).
}
\end{table}

\begin{table}
\begin{center}
\begin{tabular}{ c|ccc}
 \hline
$B_{IV}\Big(\q_i,\q_j\Big)$
& $\q_i=\sigma^x$
& $\q_i=\sigma^y$
& $\q_i=\sigma^z$ \\
\hline
 $\q_j=\sigma^x$
 & $\sin(2d)\cos(2d)$
 & $\cos^2(2d)$
 & $-\sin^2(2d)$ \\
 $\q_j=\sigma^y$
 & $-\sin^2(2d)$
 & $-\sin(2d)\cos(2d)$
 & 0 \\
 $\q_j=\sigma^z$
 & $\cos^2(2d)$
 & 0
 & $-\sin(2d)\cos(2d)$ \\
 \hline
\end{tabular}
\end{center}
\caption{\label{Table3}\small
Values of the prefactor $B_{IV}(\q_i,\q_j)$ entering
eq.~(\ref{BIIIdef}).
}
\end{table}

The contribution associated with the eigenvector pair
$\E_3^{(L,R)}$ decays with the eigenvalue
$
\lambda_3=\cos^2(2d)\cos^2(2G)
$
and has the form
\begin{equation}\label{CexactUniformRight3}
C(t)=
\cos^2(2G)
B_{III}\Big(\q_1,\q_2,\q_3\Big)
B_{III}\Big(\q_8,\q_7,\q_6\Big)
\lambda_3^{t-2}.
\end{equation}
Here,
\begin{equation}\label{BIIIdef}
B_{III}\Big(\q_i,\q_j,\q_k\Big)=
\left\{
\begin{array}{ccl}
\sin(2G) B_{IV}\Big(\q_i,\q_j\Big),
&& \q_k=\sigma^0,\\[1mm]
\cos(2G) A_{II}\Big(\q_i,\q_j\Big),
&& \q_k=\sigma^z,\;\sigma^y.
\end{array}
\right.
\end{equation}
The values of $A_{II}(\q_i,\q_j)$ are given in
Table~\ref{Table11}, while the values of
$B_{IV}(\q_i,\q_j)$ are listed in Table~\ref{Table3}.
This contribution is nonzero for the following set of local
operators:
\begin{itemize}
    \item $\q_4=\sigma^z$, $\q_5=\sigma^y$, with
    $\q_1=\set{\sigma^x,\sigma^y}$,
    $\q_2=\set{\sigma^x,\sigma^y}$,
    $\q_3=\set{\sigma^0,\sigma^z}$,
    $\q_6=\set{\sigma^0,\sigma^y}$,
    $\q_7=\set{\sigma^x,\sigma^z}$, and
    $\q_8=\set{\sigma^x,\sigma^z}$.
\end{itemize}

\end{appendix}

\bibliography{references_kicLim_Corr1}

\end{document}